\documentclass[12pt]{article}
\usepackage{a4wide}
\usepackage{amsmath}
\usepackage{amssymb}
\usepackage{amsopn}
\begin{document}
{\renewcommand{\thefootnote}{\fnsymbol{footnote}}
\hfill  AEI--2005--171, NI05065\\ 
\medskip
\hfill gr--qc/0511108\\
\medskip
\begin{center}
{\LARGE  Spherically Symmetric Quantum Geometry:\\ Hamiltonian Constraint}\\
\vspace{1.5em}
Martin Bojowald\footnote{e-mail address: {\tt mabo@aei.mpg.de}}
and Rafal Swiderski\footnote{e-mail address: {\tt swidersk@maths.ox.ac.uk}, new address: Mathematical Institute, University of Oxford, 24--29 St Giles',
Oxford OX1 3LB,
United Kingdom}
\\
\vspace{0.5em}
Max-Planck-Institut f\"ur Gravitationsphysik, Albert-Einstein-Institut,\\
Am M\"uhlenberg 1, D-14476 Potsdam, Germany
\vspace{1.5em}
\end{center}
}

\setcounter{footnote}{0}

\newcommand{\Lam}{\Lambda}
\newcommand{\vt}{\vartheta}
\newcommand{\vp}{\varphi}
\newcommand{\abs}[1]{\lvert#1\rvert}

\newcommand{\case}[2]{{\textstyle \frac{#1}{#2}}}
\newcommand{\lP}{\ell_{\mathrm P}}

\newcommand{\md}{{\mathrm{d}}}
\newcommand{\Kern}{\mathop{\mathrm{ker}}}
\newcommand{\tr}{\mathop{\mathrm{tr}}}
\newcommand{\sgn}{\mathop{\mathrm{sgn}}\nolimits}

\newcommand*{\R}{{\mathbb R}}
\newcommand*{\N}{{\mathbb N}}
\newcommand*{\Z}{{\mathbb Z}}
\newcommand*{\Q}{{\mathbb Q}}
\newcommand*{\C}{{\mathbb C}}

\begin{abstract}
 Variables adapted to the quantum dynamics of spherically symmetric
 models are introduced, which further simplify the spherically
 symmetric volume operator and allow an explicit computation of all
 matrix elements of the Euclidean and Lorentzian Hamiltonian
 constraints. The construction fits completely into the general
 scheme available in loop quantum gravity for the quantization of the
 full theory as well as symmetric models. This then presents a further
 consistency check of the whole scheme in inhomogeneous situations,
 lending further credence to the physical results obtained so far
 mainly in homogeneous models. New applications in particular of the
 spherically symmetric model in the context of black hole physics are
 discussed.
\end{abstract}

\section{Introduction}

Loop quantum gravity \cite{Rov:Loops,ThomasRev,ALRev} provides a
candidate for a non-perturbative, background independent quantization
of general relativity which has already led to several results
concerning the quantum structure of space and time. There are,
however, also open issues mainly in the context of understanding the
dynamics and the semiclassical limit. While dealing with these
problems in full generality is complicated, one can isolate particular
aspects by looking at reduced situations where only a select class of
degrees of freedom is considered. This class of degrees of freedom
needs to be adapted to the physical situation of interest, which is
most commonly done by employing symmetry reduction. In the case of
loop quantum gravity, or any diffeomorphism invariant quantum theory
of connections, there is a general scheme to introduce symmetries at
the level of quantum states and basic operators \cite{SymmRed}.

Homogeneous models \cite{CosmoI,IsoCosmo,Bohr,HomCosmo,Spin} in the
context of loop quantum cosmology \cite{LoopCosRev,WS:MB,LivRev} are
by now well-understood both from the dynamical point of view and
concerning semiclassical properties
\cite{SemiClass,Time,SemiClassEmerge,APS}. They have led to tests of
and new insigths into the full theory, and resulted in many physical
applications
\cite{Sing,DynIn,Inflation,InflationWMAP,PowerLoop,NonChaos,BounceClosed,Oscill,EmergentLoop,Collapse,BHPara}.
However, field theory aspects which play an essential role in the full
theory cannot be tested by restricting oneself to homogeneous cases
such that a generalization to inhomogeneous models is needed. The
simplest inhomogeneous model is the spherically symmetric one since it
has only one physical degree of freedom
\cite{SphKl1,SphKl2,Kuchar}. States and basic operators
\cite{SphSymm}, as well as the volume operator
\cite{SphSymmVol}, have been derived along the lines of a loop
quantization based on connection variables, resulting in explicit
expressions in particular for all volume eigenvalues. Having an
explicitly known volume spectrum, which is not available in the full
theory \cite{AreaVol,Vol2,RecTh,Vol,VolNum}, was one of the main
ingredients that resulted in direct calculations in homogeneous
models, and so one may expect similar applications of the spherically
symmetric model. In particular the Hamiltonian constraint contains the
volume operator in commutators with holonomies \cite{QSDI}, which need
to be computed in order to know the constraint equation
explicitly. However, it turned out that in the quantization of
\cite{SphSymmVol} the spherically symmetric volume operator has
eigenstates different from the triad eigenstates on which holonomies
would have a simple action. This implies that even with an explicitly
known volume spectrum commutators of the volume operator with
holonomies are hard to compute in general. As a consequence,
coefficients of the constraint equation would have a complicated form.

An additional complication for setting up the Hamiltonian constraint
is presented by the non-vanishing spin connection. As discussed in
\cite{Spin}, one often has to split off non-vanishing components of the spin
connection from holonomies along homogeneous directions in order to
ensure the correct classical limit. The main observation of the
present paper will be the fact that one can advantageously split off
the connection components already at the level of states, not only
when constructing the Hamiltonian constraint operator. We will show
that this can be done in a way consistent with both the full theory
and the treatment in homogenous models, and even leads to a
considerable simplification of the volume operator. That this is
possible depends sensitively on non-trivial properties of the spin
connection and extrinsic curvature for spherically symmetric
configurations, which also hold true in polarized cylindrical wave
models. Thus, similar constructions can be done in other models, where
the method of \cite{SphSymmVol} to quantize the volume operator and to
find its spectrum would not work or where calculations would be more
complicated \cite{CylWaveVol,PolCylVol}.

We will first discuss the spherically symmetric classical phase space
and constraints in connection variables in
Sec.~\ref{s:ClassPhase}. After computing the spin connection we will
start Sec.~\ref{s:Spin} with preliminary aspects of the classical
limit, which motivates the introduction of a canonical transformation
to variables better suited to a loop quantization and its
investigation. The loop representation is then briefly done in
Sec.~\ref{s:Rep}, before Sec.~\ref{s:Ham} presents the Hamiltonian
constraint operator as the main part of this paper. This allows
several conclusions also on general aspects of loop quantum gravity
and physical properties to be discussed in Sec.~\ref{s:Disc}

\section{Classical phase space}
\label{s:ClassPhase}

A loop quantization of gravitational systems is based on real Ashtekar
variables \cite{AshVar,AshVarReell} which are given by the
su(2)-connection $A_a^i=\Gamma_a^i+\gamma K_a^i$ and its momentum, the
densitized triad $E^a_i$. In the definition of $A$, $\Gamma_a^i$ are
spin connection components compatible with the triad, $K_a^i$ are
extrinsic curvature components and $\gamma\in\R^+$ is the
Barbero--Immirzi parameter \cite{AshVarReell,Immirzi}. In spherical
symmetry, one only considers connections and triads which are
invariant under rotations up to gauge transformations, which implies
the general form
\begin{equation} \label{A}
 A=A_x(x)\Lam_3\md x+(A_1(x)\Lam_1+A_2(x)\Lam_2)\md\vt+
(A_1(x)\Lam_2-A_2(x)\Lam_1)\sin\vt\md\vp+ \Lam_3\cos\vt\md\vp
\end{equation}
and
\begin{equation} \label{E}
 E=E^x(x)\Lam_3\sin\vt\frac{\partial}{\partial x}+
(E^1(x)\Lam_1+E^2(x)\Lam_2)\sin\vt\frac{\partial}{\partial\vt}+
(E^1(x)\Lam_2-E^2(x)\Lam_1)\frac{\partial}{\partial\vp}
\end{equation}
with real functions $A_x$, $A_1$, $A_2$, $E^x$, $E^1$ and $E^2$ on the
radial manifold $B$ coordinatized by $x$ (see, e.g.,
\cite{SymmRed,LivRev}). The ${\rm su}(2)$-matrices $\Lam_I$ are
constant and are identical to $\tau_I=-\frac{i}{2}\sigma_I$ or a rigid
rotation thereof, which can be eliminated by partially fixing the
Gauss constraint. The functions $E^x$, $E^1$ and $E^2$ on $B$ are
canonically conjugate to $A_x$, $A_1$ and $A_2$:
\begin{equation}
 \Omega_{B}=\frac{1}{2\gamma G}\int_B\md x(\md
A_x\wedge\md E^x+ 2\md A_1\wedge\md E^1+2\md A_2\wedge\md E^2)
\end{equation}
with the gravitational constant $G$.

These variables are subject to constraints which are obtained by
inserting the invariant forms into the full expressions. We have the
Gauss constraint
\begin{equation}
 G[\lambda]=\int_B\md x\lambda (E^x{}'+2A_1E^2-2A_2E^1)\approx0
\end{equation}
generating U(1)-gauge transformations, the diffeomorphism constraint
\begin{equation}
 D[N_x]=\int_B\md x N_x(2A_1'E^1+2A_2'E^2-A_xE^x{}')
\end{equation}
and the Hamiltonian constraint
\begin{eqnarray}
 H[N]&=&(2G)^{-1}\int_B\md x N \left(\abs{E^x}((E^1)^2+(E^2)^2)
\right)^{-1/2} \\
&&\times\left(2E^x(E^1A_2'-E^2A_1')+2A_xE^x(A_1E^1+A_2E^2)+(A_1^2+A_2^2-1)
((E^1)^2+(E^2)^2)\right.\nonumber\\
&&\quad -(1+\gamma^2)\left.\left(2K_xE^x(K_1E^1+K_2E^2)+(K_1^2+K_2^2)((E^1)^2+
(E^2)^2)\right)\right) \nonumber\\
 &=:& -H^{\rm E}[N]+P[N]
\end{eqnarray}
where $H^{\rm E}$ is the first (so-called Euclidean) part depending
explicitly on connection components and $P$ the second part depending
on extrinsic curvature components (which are themselves functions of
$A_a^i$ and $E_i^a$).

In \cite{SphSymm} variables
\begin{eqnarray}
 A_{\vp}(x) &:=& \sqrt{A_1(x)^2+A_2(x)^2}\,,\\ \label{Avp}
 E^{\vp}(x) &:=& \sqrt{E^1(x)^2+E^2(x)^2} \label{Evp}
\end{eqnarray}
and $\alpha(x)$, $\beta(x)$ defined by
\begin{eqnarray}
 \Lambda_{\vp}^A(x) &=:& \Lam_1\cos\beta(x)+\Lam_2\sin\beta(x)\,,\\
 \Lambda^{\vp}_E(x) &=:& \Lam_1\cos\left(\alpha(x)+\beta(x)\right)+
\Lam_2\sin\left(\alpha(x)+\beta(x)\right) \,.
\end{eqnarray}
for the internal directions
\begin{eqnarray}
 \Lambda_{\vp}^A(x) &:=& (A_1(x)\Lam_2-A_2(x)\Lam_1)/A_{\vp}(x)\,,\\
 \Lambda^{\vp}_E(x) &:=& (E^1(x)\Lam_2-E^2(x)\Lam_1)/E^{\vp}(x)
\end{eqnarray}
were introduced. Similarly, we have $\Lambda_{\vt}^A(x)=
-\Lam_1\sin\beta(x)+ \Lam_2\cos\beta(x)$ and the analogous expression
for $\Lambda_E^{\vt}$.  These variables are adapted to a loop
quantization in that holonomies along integral curves of generators of
the symmetry group are of the form $\exp A_{\vp}\Lambda_{\vp}^A$.

However, $E^{\vp}$ is {\em not} the momentum conjugate to $A_{\vp}$,
which instead is given by
\begin{equation} \label{Pphi}
 P^{\vp}(x):=2E^{\vp}(x)\cos\alpha(x)\,.
\end{equation}
Canonical coordinates are thus the conjugate pairs
$A_x,E^x;A_{\vp},P^{\vp};\beta,P^{\beta}$ with
\begin{equation} \label{Pbeta}
 P^{\beta}(x):=2A_{\vp}(x)E^{\vp}(x)\sin\alpha(x)=
A_{\vp}(x)P^{\vp}(x)\tan\alpha(x)
\end{equation}
The momenta as basic variables will directly be quantized, resulting
in flux operators with equidistant discrete spectra. But unlike in the
full theory and homogeneous models, the resulting quantum
representation has a volume operator which does not commute with flux
operators. This follows since volume is determined by triad
components, in particular $E^{\vp}$ which is related to $P^{\vp}$ in a
rather complicated way {\em involving the connection component
$A_{\vp}$}. Thus, the volume operator has eigenstates different from
flux eigenstates, which makes the computation of commutators with
holonomies more complicated \cite{SphSymmVol}. An alternative way to
derive volume eigenvalues in a quantization based on the variables
$A_x$, $A_1$ and $A_2$ is being pursued in
\cite{CylWaveVol,PolCylVol}.

In this paper we will considerably simplify the formalism by applying
a canonical transformation such that now $E^{\vp}$ plays the role of a
basic momentum variable. This will, of course, change the
configuration variables which will no longer be purely connection
components. At first sight, this seems to render the procedure
unsuitable for a loop quantization where basic operators make
use of holonomies of the connection. After a transformation of the
canonical variables, holonomies in general will be complicated
functions of the new variables such that the new quantum
representation would not be suitable for a loop quantization. It turns
out, however, that the special form of a spherically symmetric spin
connection and extrinsic curvature for a given triad leads to new
variables which are ideally suited to a loop representation even from
the dynamical point of view.

\section{The role of the spin connection}
\label{s:Spin}

The co-triad corresponding to a densitized triad (\ref{E}) is given by
\begin{equation}
 e = e_x\Lam_3\md x+ e_{\vp}\Lambda_E^{\vt}\md\vt+
 e_{\vp}\Lambda_E^{\vp}\sin\vt\md\vp
\end{equation}
with
\begin{equation}
 e_{\vp}=\sqrt{|E^x|} \quad\mbox{ and }\quad e_x = \sgn(E^x)
 \frac{E^{\vp}}{\sqrt{|E^x|}}\,.
\end{equation}
From this form one can compute the spin connection
\begin{equation}\label{Gamma}
 \Gamma = -(\alpha+\beta)'\Lam^3\md x+
 \frac{e_{\vp}'}{e_x}\Lambda_E^{\vp} \md\vt-
 \frac{e_{\vp}'}{e_x}\Lambda_E^{\vt} \sin\vt\md\vp+ \Lam_3\cos\vt\md\vp
\end{equation}
and the extrinsic curvature for lapse function $N$ and shift $N^x$,
\begin{equation} \label{K}
 K=N^{-1}(\dot{e}_x-(N^xe_x)')\Lam_3\md x+
 N^{-1}(\dot{e}_{\vp}-N^xe_{\vp}')\Lambda_E^{\vt}\md\vt+
 N^{-1}(\dot{e}_{\vp}-N^xe_{\vp}')\Lambda_E^{\vp}\sin\vt\md\vp\,.
\end{equation}
We define the $\vp$-components of $\Gamma$ and $K$ as
\begin{equation} \label{GammaK}
 \Gamma_{\vp}:=-\frac{e_{\vp}'}{e_x} = -\frac{E^x{}'}{2E^{\vp}} \quad,\quad
 K_{\vp}:=N^{-1}(\dot{e}_{\vp}-N^xe_{\vp}')
\end{equation}
which combines to $A_{\vp}=\sqrt{\Gamma_{\vp}^2+\gamma^2K_{\vp}^2}$
since the two internal $\vp$-directions are orthogonal.

\subsection{Classical limit}
\label{s:GammaClass}

The form of spin connection and extrinsic curvature is important to
understand classical regimes in which extrinsic curvature is
small. Since the spin connection in general is not a tensor, a similar
statement about the smallness of $\Gamma$ would not have invariant
meaning since one can always choose local coordinates such that the
components $\Gamma_a^i$ are arbitrarily small. However, in a symmetric
model this may not be true for all components because within the model
only coordinate transformations preserving the invariant form
(\ref{A}), (\ref{E}) are allowed. Then, some components of the spin
connection can become covariant such that statements about their
magnitude become meaningful. In the above form, for instance, it is
clear that the $x$-component of the spin connection is not of definite
magnitude since locally one can simply gauge the angle $\alpha+\beta$
to be constant. The last term in (\ref{Gamma}), on the other hand,
does not depend on the fields at all and is thus always of the order
one. In general, components of the spin connection in inhomogeneous
directions, just as all components in the full theory, do not have
invariant meaning, while components along symmetry orbits are gauge
invariant. (This follows easily from the transformation properties of
a connection where the inhomogeneous part $g^{-1}\md g$ only
contributes to inhomogeneous directions such as $x$ when the gauge
transformation $g$ is required to preserve the symmetry and is thus
constant along symmetry orbits.) Moreover, $\Gamma_{\vp}$ is covariant
and transforms as a scalar because both $E^{x\prime}$ and $E^{\vp}$
are densities.

Covariant components of the spin connection carry information about
the intrinsic curvature of symmetry orbits. In contrast to extrinsic
curvature, this is not necessarily small in classical regimes as can
be seen for the spin connection (\ref{Gamma}) specialized to the
Schwarzschild solution. Inserting $e_{\vp}=x$ and
$e_x=(1-2M/x)^{-1/2}$ leads to $\Gamma_{\vp}=1+O(x^{-1})$ which is not
small at large radii where classical properties are to be
expected. Thus, also the Ashtekar connection component
$A_{\vp}=\sqrt{\Gamma_{\vp}^2+\gamma^2K_{\vp}^2}$ is not small in this
regime but of the order one. Consequently, angular holonomies of the
form $\exp A_{\vp}\Lambda^A_{\vp}$ cannot be expanded in $A_{\vp}$ for
semiclassical considerations. It is then more complicated to quantize
classical expressions which are polynomial in $A_{\vp}$, a prominent
example being the Hamiltonian constraint, because they first need to
be expresses through holonomies such that the classical expression is
reproduced in semiclassical regimes.

One can arrive at expressions expandable in connection components if
one explicitly subtracts the spin connection for homogeneous
directions. Instead of working with holonomies of $A_{\vp}$ one would
then construct operators from holonomies of, in the above case,
$A_{\vp}-1$ which would be small. This procedure can in fact be
described as a general scheme which works in all homogenous models
and gives a satisfactory evolution there (i.e.\ stable in the sense of
\cite{FundamentalDisc} and with the correct classical limit).

The same procedure can be followed in the spherically symmetric model
for holonomies along symmetry orbits. However, in the variables
described so far the same complication as with the volume operator
arises: In order to subtract the angular spin connection components at
the operator level we need to quantize the spin connection which,
since it is related to triad components, results in a complicated
expression in terms of canonical variables $A_{\vp}$ and $P^{\vp}$.

\subsection{Canonical triad variables}

The preceding remarks have shown that many steps in the usual
constructions of a loop quantization become much more complicated when
triad components are not among the basic canonical variables. On the
other hand, applying a canonical transformation such that $P^{\vp}$ is
replaced by $E^{\vp}$ may lead to more complicated configuration
variables which are no longer related to holonomies in a simple
way. We will now see that the explicit form (\ref{Gamma}) of the spin
connection and (\ref{K}) of extrinsic curvature in spherical symmetry
allows one to perform a suitable canonical transformation and at the
same time facilitate the subtraction of spin connection components
already at the state level. The subtraction in the constraint operator
can then be done easily. There is a difference to the treatment of
homogeneous models in \cite{Spin} where the spin connection was used
explicitly only at the operator level. However, there the subtraction
could also have been done at the state level such that the procedure
here is consistent with \cite{Spin}. Moreover, we will see that
inhomogeneous directions are not affected by the subtraction and we
are also consistent with the full theory. We thus have crucial tests
available by models which are in between homogeneous ones and the full
theory, as discussed in more detail below after explicit constructions
will be available.

Since the momentum of $A_x$ is already given by a triad component
$E^x$, it will be unchanged by our canonical transformation and we can
focus on the variables $A_{\vp},P^{\vp};\beta,P^{\beta}$. Using the
definitions (\ref{Pphi}) and (\ref{Pbeta}) of $P^{\vp}$ and
$P^{\beta}$ in the canonical Liouville form and trading in $E^{\vp}$
for $P^{\vp}$ results in
\begin{eqnarray}
 P^{\vp}\md A_{\vp}+P^{\beta}\md\beta &=& 2E^{\vp}\cos\alpha\md
 A_{\vp} + P^{\beta}\md\beta\nonumber \\
 &=& E^{\vp}\md(2A_{\vp}\cos\alpha)-2E^{\vp}A_{\vp}\md\cos\alpha+
 P^{\beta}\md\beta\nonumber\\
 &=& E^{\vp}\md(2A_{\vp}\cos\alpha)+P^{\eta}\md\eta\,. \label{symp}
\end{eqnarray}
In the last line we now have $E^{\vp}$ as momentum of the
configuration variable $2A_{\vp}\cos\alpha$, and the old
$P^{\eta}=P^{\beta}$ as momentum of the angle $\eta:=\alpha+\beta$
determining the internal triad direction.

As a function of the original variables, $A_{\vp}\cos\alpha$ looks
complicated and does not seem to be related to holonomies. In fact,
$\alpha$ is a function of both $A_{\vp}$ and the momenta $P^{\vp}$ and
$P^{\beta}$ such that it cannot be expressed as a function of
holonomies in the original variables alone. However, the structure of
(\ref{Gamma}) and (\ref{K}) shows that there is a simple geometrical
meaning to the new configuration variable conjugate to $E^{\vp}$. Here
it is important to notice that the internal directions along a given
angular direction of a spherically symmetric $\Gamma$ in (\ref{Gamma})
are always perpendicular to those of $E$ (note that $\Lambda_E^{\vt}$
and $\Lambda_E^{\vp}$ are exchanged in (\ref{Gamma}) compared to
(\ref{E})), while the corresponding extrinsic curvature components are
parallel to those of $E$. Since $A$ is obtained by summing $\Gamma$
and $K$, we can write
\[
 A_{\vp}\Lambda_{\vp}^A = \Gamma_{\vp} \bar\Lambda +\gamma
 K_{\vp}\Lambda
\]
with $\Lambda:=\Lambda_E^{\vp}$ and
$\bar\Lambda:=\Lambda_E^{\vt}$. This implies
\begin{equation}
 A_{\vp}\cos\alpha = A_{\vp}\Lambda_{\vp}^A\cdot\Lambda =
 \gamma K_{\vp}
\end{equation}
where the left equality is the definition of $\alpha$. Thus, the new
configuration variable is simply proportional to the extrinsic
curvature component $K_{\vp}$ which we can view as obtained by
subtracting the spin connection from the Ashtekar connection. Note
that it is well known in the full theory that extrinsic curvature
components are conjugate to triad components. But as we have seen for
Ashtekar variables, this does not imply that the $\vp$-components as
defined here are conjugate (while $E^1$, $E^2$ would obviously be
conjugate to $A_1$, $A_2$ as well as $K_1$, $K_2$). The non-trivial
fact is thus that in contrast to the angular Ashtekar components, the
angular extrinsic curvature component as configuration variable allows
one to use triad components as momenta. As the derivation shows, this
depends crucially on properties of the spherically symmetric spin
connection and extrinsic curvature. That $E^{\vp}$ is conjugate to
$K_{\vp}$ follows from the fact that $E$ and $K$ have the same internal
directions, while the orthogonality of internal directions in $\Gamma$
to those of $E$ is relevant for details of the canonical transformation.

\subsection{Hamiltonian constraint}

In the original spherically symmetric variables the Hamiltonian
constraint has Euclidean part
\[
 H^{\rm E}[N] = -(2G)^{-1}\int_B\md x N(x) |E^x|^{-1/2}\left(
 (A_{\vp}^2-1)E^{\vp}+2\cos\alpha A_{\vp}E^x(A_x+\beta')- 2\sin\alpha
 E^x A_{\vp}'\right)\,.
\]
With the new canonical triad variables, using
\[
  A_{\vp}'\sin\alpha = ( A_{\vp}\sin\alpha)'- A_{\vp}\alpha'\cos\alpha
 = \Gamma_{\vp}'-\gamma K_{\vp}\alpha'
\]
we have
\begin{equation} \label{HE}
 H^{\rm E}[N] = -(2G)^{-1}\int_B\md x N(x) |E^x|^{-1/2}\left(
 (\Gamma_{\vp}^2+\gamma^2K_{\vp}^2-1)E^{\vp}+2\gamma
 K_{\vp}E^x(A_x+\eta')- 2E^x\Gamma_{\vp}' \right)\,.
\end{equation}
Moreover, the Lorentzian part is simply of the form
\[
 P[N]=-(2G)^{-1}(1+\gamma^2)\int_B\md x N(x) |E^x|^{-1/2}\left(
K_{\vp}^2E^{\vp}+2K_{\vp}K_xE^x \right)
\]
which, using $\gamma K_x=A_x-\Gamma_x = A_x+\eta'$,
combines easily with the terms already present in (\ref{HE}):
\begin{equation} \label{HL}
 H[N] = -(2G)^{-1}\int_B\md x N(x) |E^x|^{-1/2}\left(
 (1-\Gamma_{\vp}^2+K_{\vp}^2)E^{\vp}+2\gamma^{-1}
 K_{\vp}E^x(A_x+\eta')+ 2E^x\Gamma_{\vp}' \right)  \,.
\end{equation}

\section{Loop representation}
\label{s:Rep}

With the new spherically symmetric configuration variables
$A_x,\gamma K_{\vp},\eta$ the construction of the quantum
representation proceeds identically to that in \cite{SphSymm}, just by
replacing everywhere $A_{\vp}$ with $\gamma K_{\vp}$ and $\beta$ with
$\eta$. The resulting Hilbert space is then spanned by an
orthonormal basis in terms of spin network states
\begin{equation} \label{SpinNetwork}
 T_{g,k,\mu}(A) = \prod_{e\in g}
 \exp\left(\tfrac{1}{2}ik_e\smallint_e A_x(x)\md x\right) \prod_{v\in
 V(g)} \exp(i\mu_v \gamma K_{\vp}(v)) \exp(ik_v\eta(v))
\end{equation}
with edge labels $k_e\in\Z$ and vertex labels $\mu_v\in\R$ and
$k_v\in\Z$ for graphs $g$ in the 1-dimensional radial manifold $B$.

By definition in (\ref{Avp}), $A_{\vp}$ is always non-negative which
is sufficient because a sign change in both components $A_1$ and $A_2$
can always be compensated by a gauge rotation. The extrinsic curvature
component $K_{\vp}$, on the other hand, is measured relatively to the
internal direction $\Lambda_E^{\vp}$ in (\ref{K}) and thus both signs
are possible: $K_{\vp}\in{\mathbb R}$. This makes the representation
technically easier to deal with, but could be done similarly with
connection components upon defining $\bar{A}_{\vp}:=A_{\vp}\sgn
K_{\vp}$. (See also \cite{BHInt} for a discussion in a homogeneous
model.)

Similarly, we immediately obtain the action of flux operators
quantizing the momenta $E^x$, $E^{\vp}$ and $P^{\eta}$ (using the
Planck length $\lP=\sqrt{G\hbar}$):
\begin{eqnarray}
 \hat{E}^x(x) T_{g,k,\mu} &=& \gamma\lP^2
\frac{k_{e^+(x)}+k_{e^-(x)}}{2} T_{g,k,\mu} \label{Exspec}\\
 \int_{\cal I}\hat{E}^{\vp}T_{g,k,\mu} &=& \gamma\lP^2
\sum_{v\in{\cal I}} \mu_v T_{g,k,\mu}\label{Epspec}\\
 \int_{\cal I}\hat{P}^{\eta} T_{g,k,\mu} &=&
2\gamma\lP^2\sum_{v\in{\cal I}} k_v T_{g,k,\mu} \label{Pbspec}
\end{eqnarray}
where $e^{\pm}(x)$ are the two edges (or two parts of a single edge)
meeting in $x$. (Compared to the expression for $\hat{P}^{\vp}$ in
\cite{SphSymm} there is a factor of $\frac{1}{2}$ missing in
(\ref{Epspec}) since $E^{\vp}$ is conjugate to $\gamma K_{\vp}/G$,
while $P^{\vp}$ was defined to be conjugate to $A_{\vp}/2G$; see
(\ref{symp}). Note also that in some previous papers $\ell_{\rm
P}^2=8\pi G\hbar$ was used.)  Spin networks, as expected, are thus
eigenstates of the flux operators with the crucial difference to
\cite{SphSymm} being that now $\hat{E}^{\vp}$ is among the fluxes.

This allows us to quantize the
volume $V({\cal I})=4\pi \int_{{\cal I}}\md x
\sqrt{|E^x|}E^{\vp}$ of a region ${\cal I}\subset B$ immediately after
regularizing as in \cite{SphSymmVol}, resulting in the volume operator
\begin{equation}
 \hat{V}({\cal I}) = 4\pi \int_{\cal I} \md x|\hat{E}^{\vp}(x)|
\sqrt{|\hat{E}^x(x)|} 
\end{equation}
where $\hat{E}^{\vp}(x)$ is the distribution valued operator
\[
 \hat{E}^{\vp}(x)T_{g,k,\mu} = \gamma\lP^2
  \sum_{v\in B} \delta(v,x) \mu_v T_{g,k,\mu}\,.
\]
Note that, just as $A_{\vp}$ in (\ref{Avp}), also $E^{\vp}$ is defined
to be non-negative in (\ref{Evp}). Thus, only labels $\mu_v\geq 0$
would be allowed. Again, it is technically easier to allow first all
values $\mu_v\in{\mathbb R}$ and in the end require physical states to
be symmetric under $\mu_v\mapsto-\mu_v$ corresponding to solving a
residual gauge transformation. We thus write explicitly absolute
values around $\hat{E}^{\vp}(x)$ and $\mu_v$.  The volume operator
then has eigenstates (\ref{SpinNetwork}) with eigenvalues
\begin{equation}
 V_{k,\mu} = 4\pi\gamma^{3/2}\lP^3 \sum_v
 |\mu_v|\sqrt{\case{1}{2}|k_{e^+(v)}+k_{e^-(v)}|} \,.
\end{equation}
In contrast to \cite{SphSymmVol}, the eigenvalues follow immediately
and the eigenstates are identical to flux eigenstates.

The Gauss constraint and the fluxes it depends on are unmodified as
compared to \cite{SphSymm} (except for the $-$ sign in front of
$P^{\beta}$ which was written by mistake in \cite{SphSymm}) such that
classically it still is given by
\begin{equation} \label{Gauss}
 G[\lambda]=\int_B\md x\lambda(E^x{}'+P^{\eta})\approx0
\end{equation}
and quantized to
\begin{equation}
 \hat{G}[\lambda] T_{g,k,\mu} = \gamma\lP^2\sum_v \lambda(v)
(k_{e^+(v)}-k_{e^-(v)}+2k_v) T_{g,k,\mu} = 0\,.
\end{equation}
As before, this is solved directly and imposes the condition
\begin{equation} \label{kGauss}
 k_v=-\tfrac{1}{2}(k_{e^+(v)}-k_{e^-(v)})
\end{equation}
on gauge invariant states. Inserting this in (\ref{SpinNetwork}) we
eliminate the integer valued vertex labels $k_v$ and obtain the
general form of gauge invariant spherically symmetric spin networks
\begin{equation} \label{GaugeInvSpinNetwork}
 T_{g,k,\mu}=\prod_e \exp\left(\tfrac{1}{2}i k_e
\smallint_e(A_x+\eta')\md x\right)  \prod_v
\exp(i\mu_v \gamma K_{\vp}(v))\,.
\end{equation}
They depend only on the gauge invariant configuration
variables $A_x+\eta'=\gamma K_x$ and $K_{\vp}$.

\section{Hamiltonian constraint}
\label{s:Ham}

There is a general procedure to quantize objects such as the Euclidean
part
\[
 H^{\rm E}[N] = -(8\pi G)^{-1}\int_{\Sigma}\md^3xN(x)\epsilon^{ijk} F_{ab}^i
 \frac{E^a_jE^b_k}{\sqrt{|\det E|}}
\]
of the Hamiltonian constraint in the full theory, where the curvature
components $F_{ab}^i(x)$ of the Ashtekar connection are expressed via
a holonomy around a small loop starting at $x$, and the factor
containing triad components is obtained from a commutator between
holonomies and the volume operator
\cite{QSDI}. This procedure can also be adapted to symmetric models,
which allows detailed tests of its viability on general grounds. While
the commutator part goes through almost unchanged in symmetric models
(see, however, \cite{DegFull}), the quantization of curvature
components is different. For one obtains the correct components only
in an expansion of the holonomy for a closed loop, which implies that
the exponent appearing in this holonomy must be small. This is easily
achieved in the full theory by making the loops small enough in
coordinate length, which results automatically in the continuum limit
in which a regulator is removed. A similar argument can be applied to
inhomogeneous directions in symmetric models such as the radial
direction in the system studied here. We can then use, e.g.,
\[
 \exp i\int_v^{v_+} A_x \md x\sim 1+i\epsilon A_x(v)+O(\epsilon^2)
\]
when $\epsilon=v_+-v$ is the coordinate distance between two vertices
$v$ and $v_+$ connected by the edge.

For homogeneous directions, however, this argument cannot be applied
since there is no free edge length available which could be chosen
small. Homogeneous connection components transform as scalars and thus
are implemented via `point holonomies'
\cite{FermionHiggs,ScalarBohr,SymmRed,Bohr} such as $\exp i\gamma
K_{\vp}(v)$ rather than holonomies along edges. An expansion
\[
 \exp iA(v) \sim 1+iA(v)+O(A^2)
\]
is then possible only in regimes where the relevant connection
component $A$ is small. (Alternatively, one can employ a limit
$\gamma\to0$ as in \cite{SemiClass}, but it must still be shown to
exist in an inhomogeneous construction. This may also require
conditions on the basic algebra used.) Since the expansions are done
in order to reproduce the classical limit, only exponents in
holonomies are allowed which become small in semiclassical regimes. In
general, there will be other regimes in which the components are not
small, which will give rise to perturbative quantum corrections
\cite{EffHam,DiscCorr,AmbigConstr,SemiClassEmerge,Perturb,AnisoPert}.

\subsection{Classical limit}

Smallness in semiclassical regimes is not guaranteed for $A_{\vp}$, as
discussed before in Sec.~\ref{s:GammaClass}, and so one would need to
subtract off the spin connection from this component along the lines
of \cite{Spin} in order to ensure the correct classical limit. This is
complicated in the spherically symmetric model since the spin
connection, containing $E^{\vp}$, would be a complicated operator in a
loop quantization based on configuration variables containing
$A_{\vp}$. In the quantization described here, $\hat{E}^{\vp}$ and
thus $\hat{\Gamma}_{\vp}$ are simpler, but a subtraction is not even
necessary since now $K_{\vp}$ is a basic configuration variable. Since
the extrinsic curvature component is small in semiclassical regimes,
we can work directly with holonomies $\exp i\gamma K_{\vp}$ which are
basic and also appear in spin network states (\ref{SpinNetwork}).

When there is non-vanishing intrinsic curvature in a model, one can
proceed by splitting the Hamiltonian constraint into two parts which
will be quantized separately \cite{Spin}. The
first one just contains extrinsic curvature components besides triad
components and will be quantized by employing holonomies around
suitable loops. In our case, this part is classically given by the
contribution
\begin{equation} \label{HK}
 H_K[N] = -(2G)^{-1}\int_B\md x N(x) |E^x|^{-1/2}\left(
 K_{\vp}^2 E^{\vp}+2\gamma^{-1}
 K_{\vp}(A_x+\eta') E^x) \right)
\end{equation}
to (\ref{HL}). The second part contains the curvature of the spin
connection, which in our case results in
\begin{equation} \label{HG}
 H_{\Gamma}[N] = -(2G)^{-1}\int_B\md x N(x) |E^x|^{-1/2}\left(
 (1-\Gamma_{\vp}^2)E^{\vp}+ 2\Gamma_{\vp}' E^x \right)\,.
\end{equation}

Both parts consist of two terms containing $E^{\vp}$ and $E^x$,
respectively. Via $E^{\vp}/\sqrt{|E^x|}\propto \{A_x,V\}$ and
$\sqrt{|E^x|}\propto \{K_{\vp},V\}$ these terms will contain
commutators $h_x[h_x^{-1},\hat{V}]$ and
$h_{\vp}[h_{\vp}^{-1},\hat{V}]$, respectively, when quantized. Here
and in what follows we use holonomies
\begin{eqnarray}
 h_x(e) &=& \exp(\smallint_e A_x(x)\md x\Lam_3)\\
 h_{\vt}(v,\delta) &=& \exp(\gamma\delta K_{\vp}(v)\bar\Lambda)\\
 h_{\vp}(v,\delta) &=& \exp(\gamma\delta K_{\vp}(v)\Lambda)
\end{eqnarray}
adapted to the symmetry in order to construct the operator. The edge
length and the parameter $\delta$ have to be chosen for the
construction; their roles will become clear later on.

The spin connection part $H_{\Gamma}$ will be completed by using the
expression (\ref{GammaK}) for $\Gamma_{\vp}$ and the triad
operators. Since classically the inverse of $E^{\vp}$ appears, one has
to use techniques as in \cite{QSDV,InvScale} in order to obtain a
densely defined operator. For the extrinsic curvature part $H_K$ we
need to construct appropriate products of holonomies in order to model
loops resulting in the correct curvature components. Suitable products
can be read off from the general form \cite{QSDI}
\begin{equation} \label{FullHam}
 \hat{H}\propto \sum_{v,i,j,k} N(v) \epsilon^{ijk}\tr(h_{ij}
 h_k[h_k^{-1},\hat{V}])
\end{equation}
for a quantization, where we sum over vertices $v$ and triples $i,j,k$
of edges and use the lapse function $N$ smearing the
constraint. Holonomies $h_K$ for each edge appear in the commutators,
multiplied with a loop holonomy $h_{ij}$ lying in a plane spanned by
the two other edges of the triple. In a symmetric context, this scheme
takes the form \cite{CosmoIII,IsoCosmo,HomCosmo,Spin}
\[
 \hat{H}\propto \sum_{v,I,J,K} N(v)\epsilon^{IJK}\tr(h_Ih_Jh_I^{-1}h_J^{-1}
 h_K[h_K^{-1},\hat{V}])
\]
where $h_{ij}$ is constructed from holonomies available in the
symmetric model, where now $I,J,K$ correspond to coordinates adapted
to the symmetry. In general, each holonomy $h_I$ for a homogeneous
direction appears with a parameter $\delta$ (which we will use but not
specify in what follows) with $\delta^2$ being analogous to the size
of the loop in $h_{ij}$. This parameter allows one to do formal
expansions of functions of holonomies, and to check if the correct
classical limit results to leading order, but this limit will also be
valid for small $K_{\vp}$, as expected in semiclassical regimes, even
if $\delta$ is of the order one.

In spherical symmetry, $\{I,J,K\}$ is $\{x,\vt,\vp\}$ such that we have
two essentially different contributions for $K=x$ and $K=\vt,\vp$,
respectively. They lead to operators
\[
 \tr(h_{\vt}(v)h_{\vp}(v)h_{\vt}(v)^{-1}h_{\vp}(v)^{-1}
h_x[h_x^{-1},\hat{V}])
\]
and
\[
 \tr(h_xh_{\vt}(v_+)h_x^{-1}h_{\vt}(v)^{-1}
h_{\vp}(v)[h_{\vp}(v)^{-1},\hat{V}])
\]
where the $x$-holonomy is computed for an edge connecting the vertex
$v$ to a new one, $v_+$. Using $x$-holonomies for small edge length
$\epsilon$ and expanding in $\delta$ (or using small $K_{\vp}$ in a
semiclassical regime) we see that both terms indeed give us the
expected curvature components. We first expand the product of
holonomies
\begin{eqnarray*}
 h_{\vt}h_{\vp}h_{\vt}^{-1}h_{\vp}^{-1} &=& \exp(\gamma\delta
 K_{\vp}\bar\Lambda) \exp(\gamma\delta K_{\vp}\Lambda)
 \exp(-\gamma\delta K_{\vp}\bar\Lambda) \exp(- \gamma\delta
 K_{\vp}\Lambda)\\ & = & (1+ \gamma\delta K_{\vp}\bar\Lambda+O(\delta^2))
 (1+\gamma\delta K_{\vp}\Lambda+O(\delta^2))\\ &&\times (1-\gamma\delta
 K_{\vp}\bar\Lambda+O(\delta^2)) (1-\gamma\delta K_{\vp}\Lambda+O(\delta^2))\\
 &=& 1+\case{1}{2}\gamma^2\delta^2K_{\vp}^2 +
 \gamma^2\delta^2K_{\vp}^2\Lam_3+O(\delta^3)
\end{eqnarray*}
and
\begin{eqnarray*}
 h_xh_{\vt}(v_+)h_x^{-1}h_{\vt}(v)^{-1} &=& e^{\int A_x\Lam_3}
 \exp(\gamma\delta K_{\vp}(v_+)\bar\Lambda(v_+)) e^{-\int A_x\Lam_3} 
\exp(-\gamma\delta K_{\vp}(v)\bar\Lambda(v))\\ 
 & = & (1+\smallint A_x\Lam_3 +O(\epsilon^2)) (1+ \gamma\delta
 K_{\vp}(v_+)\bar\Lambda(v_+)+O(\delta^2))\\ && \times (1- \smallint
 A_x\Lam_3+O(\epsilon^2)) (1-\gamma\delta
 K_{\vp}(v)\bar\Lambda(v)+O(\delta^2))\\ &=& 1 + \gamma\delta
 K_{\vp}(v_+)\bar\Lambda(v_+)- \gamma\delta K_{\vp}(v)\bar\Lambda(v)+
 \case{1}{2}\gamma\delta K_{\vp}(v_+)\smallint A_x \Lambda(v_+)\\ &&+
 \case{1}{2}\gamma\delta K_{\vp}(v)\smallint A_x \Lambda(v) -\gamma^2
 \delta^2K_{\vp}(v)K_{\vp}(v_+) \bar\Lambda(v)\bar\Lambda(v_+)\\ &&
 +O(\delta^3)+O(\epsilon^2)
\end{eqnarray*}
which will be multiplied by the commutators $h_x[h_x^{-1},\hat{V}]$
and $h_{\vp}(v)[h_{\vp}(v)^{-1},\hat{V}]$, respectively. To leading
order in $\epsilon$ and $\delta$, the first commutator is
proportional to $\Lam_3$, while the second one is proportional to
$\Lambda(v)$. Thus, the relevant traces to check the classical limit
are
\[
 -2\tr(h_{\vt}h_{\vp}h_{\vt}^{-1}h_{\vp}^{-1} \Lam_3) \sim
 \gamma^2\delta^2K_{\vp}^2
\]
and
\begin{eqnarray*}
 -2\tr(h_xh_{\vt}(v_+)h_x^{-1}h_{\vt}(v)^{-1} \Lambda(v)) &\sim&
 -2\gamma\delta K_{\vp}(v_+) \tr(\bar\Lambda(v_+)\Lambda(v))\\
 &&-\gamma\delta K_{\vp}(v_+)\smallint\!\! A_x \tr(\Lambda(v_+)\Lambda(v))-
 \gamma \delta K_{\vp}(v)\smallint\!\! A_x \tr(\Lambda(v)^2)\\ &=& \gamma\delta
 K_{\vp}(v_+)\sin\Delta\eta\\ &&+\case{1}{2}\gamma\delta K_{\vp}(v_+)
 \smallint\!\! A_x \cos\Delta\eta+ \case{1}{2}\gamma\delta K_{\vp}(v) 
 \smallint\!\! A_x\\ &=& \gamma\delta\epsilon
 K_{\vp}(v)(A_x(v)+\eta'(v))+O(\delta\epsilon^2)
\end{eqnarray*}
with $\Delta\eta=\eta(v_+)-\eta(v)$. With these expressions we obtain
the right coefficients of $E^x$ and $E^{\vp}$ in $H_K$ to ensure the
correct classical limit.

\subsection{Operator}

The preceding discussion, together with collecting numerical factors,
shows that the operator
\begin{eqnarray} \label{HOp}
 \hat{H}[N]\!\! &\!\!\!=\!\!\!&\!\! \frac{i}{2\pi G\gamma^3\delta^2\lP^2}
 \sum_{v,\sigma=\pm1} \sigma N(v)
 \tr\Bigl(\left(h_{\vt}h_{\vp}h_{\vt}^{-1}h_{\vp}^{-1}
-h_{\vp}h_{\vt}h_{\vp}^{-1}h_{\vt}^{-1}\right.\\
&&\qquad\qquad+2\gamma^2\delta^2
 (1-\hat{\Gamma}_{\vp}^2)\Lam_3\bigr)
 h_{x,\sigma}[h_{x,\sigma}^{-1},\hat{V}]\nonumber\\
 &&+
 \left(h_{x,\sigma}h_{\vt}(v_{\sigma})h_{x,\sigma}^{-1}h_{\vt}(v)^{-1}
-h_{\vt}(v)h_{x,\sigma}h_{\vt}(v_{\sigma})^{-1}h_{x,\sigma}^{-1}+
 2\gamma^2\delta\smallint_{\sigma}\hat{\Gamma}'_{\vp}\Lambda(v)\right)
 h_{\vp}[h_{\vp}^{-1},\hat{V}]\nonumber\\
 &&+\left.
 \left(h_{\vp}(v)h_{x,\sigma}h_{\vp}(v_{\sigma})^{-1}h_{x,\sigma}^{-1}-
h_{x,\sigma}h_{\vp}(v_{\sigma})h_{x,\sigma}^{-1}h_{\vp}(v)^{-1}+
 2\gamma^2\delta\smallint_{\sigma}\hat{\Gamma}'_{\vp}\bar\Lambda(v)\right)
 h_{\vt}[h_{\vt}^{-1},\hat{V}]\right) \nonumber
\end{eqnarray}
corresponds to the correct classical expression where in the continuum
limit coordinate differentials $\epsilon$ from $h_x$ and
$\int\Gamma_{\vp}'$ become the integration measure. (The sign factor
$\sigma$ denotes the orientation of the radial edge running to the
right ($+$) or left ($-$) of the vertex. If it is not specified, it is
understood to be positive without crucial changes for the other
orientation. Moreover, $\int_{\sigma}$ in front of the derivative of
the spin connection component indicates that it has to be integrated
between $v$ and $v_{\sigma}$ as a result of the discretization.)  To
evaluate the action explicitly we now use $\exp (A\Lambda) =
\cos\case{1}{2} A+2\Lambda\sin\case{1}{2} A$ for all holonomies. This
gives, for instance,
\begin{eqnarray*}
 h_{\vt}h_{\vp}h_{\vt}^{-1}h_{\vp}^{-1} &=& \cos^2
 (\case{1}{2}\gamma\delta K_{\vp})+ \sin^2(\case{1}{2}\gamma\delta
 K_{\vp})\cos(\gamma\delta K_{\vp})+\sin^2(\gamma\delta
K_{\vp})\Lam_3\\
 &&+
 4\cos(\case{1}{2}\gamma\delta K_{\vp})\sin^3 (\case{1}{2}\gamma\delta K_{\vp})
 (\Lambda_{\vt}-\Lambda_{\vp})
\end{eqnarray*}
and
\begin{eqnarray*}
 h_x[h_x^{-1},\hat{V}] &=& \hat{V}-\cos(\case{1}{2}\smallint A_x)
 \hat{V}\cos(\case{1}{2}\smallint A_x)- \sin(\case{1}{2}\smallint A_x )
 \hat{V}\sin(\case{1}{2}\smallint A_x)\\
 &&+ 2\Lam_3 \left(\cos(\case{1}{2}\smallint A_x)
 \hat{V}\sin(\case{1}{2}\smallint A_x)-\sin(\case{1}{2}\smallint A_x )
 \hat{V}\cos(\case{1}{2}\smallint A_x)\right)
\end{eqnarray*}
which, when combined with the contribution where $\vt$ and $\vp$ are
exchanged and traced, yields the term
\begin{eqnarray*}
&& -2\tr\left(\left(h_{\vt}h_{\vp}h_{\vt}^{-1}h_{\vp}^{-1}
-h_{\vp}h_{\vt}h_{\vp}^{-1}h_{\vt}^{-1}\right)
 h_{x,\sigma}[h_{x,\sigma}^{-1},\hat{V}]\right)=
2\sin^2 (\gamma\delta K_{\vp})\\
&&\qquad\qquad\times \left(\cos(\case{1}{2}\smallint A_x)
 \hat{V}\sin(\case{1}{2}\smallint A_x)-\sin(\case{1}{2}\smallint A_x )
 \hat{V}\cos(\case{1}{2}\smallint A_x)\right)
\end{eqnarray*}
as one part of the constraint operator.

Similarly, we obtain
\begin{eqnarray}
&& -2\tr\left(\left(h_x h_\vt(v_+) h_x^{-1} h_\vt(v)^{-1}-h_{\vt}(v) h_x
h_{\vt}(v)^{-1} h_x^{-1}\right) h_\vp [h_\vp^{-1},\hat{V}]\right. \nonumber\\
&&\quad+\left.\left(h_\vp(v) h_x h_\vp^{-1}(v_+) h_x^{-1}-h_x
h_\vp(v_+) h_x^{-1} h_\vp^{-1}(v)\right) h_\vt [h_\vt^{-1},\hat{V}]
\right)\nonumber\\
&& =4 \cos(\case{1}{2} \gamma\delta K_\vp(v)) \sin(\case{1}{2} 
\gamma\delta K_\vp(v_+)) \sin(\smallint A_x-
\eta(v)+\eta(v_+)) \label{holssin}\\ 
&&\quad\times \left(\cos(\case{1}{2} \gamma\delta K_\vp(v)) \hat{V}
\sin(\case{1}{2} \gamma\delta K_\vp(v))-\sin(\case{1}{2} \gamma\delta
K_\vp(v)) \hat{V} \cos(\case{1}{2}
\gamma\delta K_\vp(v))\right)\nonumber
\end{eqnarray}

The spin connection component $\Gamma_{\vp}= -\frac{E^x{}'}{2E^{\vp}}$
and its integrated spatial derivative
$\int_{\sigma}\Gamma_{\vp}'=\Gamma_{\vp}(v_{\sigma})-\Gamma_\vp(v)$ can
be quantized using flux operators, where we need to choose a
discretization for the derivatives and be careful with the inverse of
$E^{\vp}$ because the flux operators are not invertible. The latter
problem can be solved in a manner by now common in loop quantum
gravity, expressing the classical inverse as a Poisson bracket between
holonomies and only positive powers of flux operators \cite{QSDV}. The
derivatives are also straightforward to deal with because the total
expressions are scalar. We can thus write
\[
 \Gamma_{\vp}(v)= -\frac{E^x{}'}{2E^{\vp}}= -\frac{1}{4} \left(
\frac{E^x(v_+)-E^x(v)}{\int_+ E^{\vp}}- \frac{E^x(v_-)-E^x(v)}{\int_-
E^{\vp}}\right) +O(\epsilon)\,,
\]
treating the two neighboring vertices symmetrically, where now all
expressions in the numerators and denominators are scalar and can be
quantized. Note that in this manner $\hat{\Gamma}_{\vp}$ becomes
non-zero only in vertices. There are several choices involved in the
construction, choosing a quantizable form for the inverse
\cite{Ambig,ICGC} and a discretization of the derivative, but
qualitative aspects to be discussed in what follows are not affected.

\subsubsection{Action}

To write down the action on triad eigenstates explicitly it is
convenient to split the vertex contribution to the operator into three
parts, $\hat{H}_v= \hat{H}_{\rm L}+\hat{H}_{\rm C}+\hat{H}_{\rm R}$
with
\begin{equation}
 \hat{H}_{\rm R/L}= \frac{-i}{\pi G\gamma^3\delta^2\lP^2}
\cos(\case{1}{2} 
\gamma\delta K_\vp(v))
 \sin(\case{1}{2} \gamma\delta K_\vp(v_{\pm}))
 \sin\left(\smallint_v^{v_{\pm}} A_x-
\eta(v)+\eta(v_{\pm})\right) \Delta_{\vp}\hat{V}
\end{equation}
depending on $K_{\vp}$ in $v$ and $v_+$ or $v_-$, respectively
(receiving contributions from the two bottom lines in (\ref{HOp}) for
$\sigma=+$ for R and $\sigma=-$ for L)
and
\begin{equation}
 \hat{H}_{\rm C}=\frac{-i}{\pi G\gamma^3\delta^2\lP^2} 
\left(\sin^2 (\gamma\delta K_{\vp}) \Delta_x\hat{V}+
 \gamma^2\delta^2(1-\hat{\Gamma}_{\vp}^2) \Delta_x\hat{V}+
 \gamma^2\delta\hat{\Gamma}'_{\vp}\Delta_{\vp}\hat{V}\right) 
 +\hat{H}_{{\rm matter},v}
\end{equation}
depending on $K_{\vp}$ only in $v$ (with contributions from the top
lines in (\ref{HOp}) for both $\sigma=+$ and $\sigma=-$)
where
\begin{equation}
\Delta_x\hat{V}:= \cos(\case{1}{2}\smallint A_x)
 \hat{V}\sin(\case{1}{2}\smallint A_x)-\sin(\case{1}{2}\smallint A_x)
 \hat{V}\cos(\case{1}{2}\smallint A_x)
\end{equation}
and
\begin{equation}
 \Delta_{\vp}\hat{V}:=\cos(\case{1}{2} \gamma\delta K_\vp(v)) \hat{V}
 \sin(\case{1}{2}  
\gamma\delta K_\vp(v))-\sin(\case{1}{2} \gamma\delta K_\vp(v)) \hat{V}
\cos(\case{1}{2} \gamma\delta K_\vp(v))\,.
\end{equation}
Since the expression for $V$ in terms of $k_+$ and $k_-$ only depends
on the sum $k_++k_-$ and the operator $\Delta_x\hat{V}$ turns out to
be diagonal on triad eigenstates, we do not need to distinguish
between the versions integrating to $v_+$ and $v_-$, respectively, if
$v_{\pm}$ are new vertices. This is different if $v_{\pm}$ already
exist as vertices of the original graph which, however, does not
crucially change coefficients. We will thus suppress this additional
possibility in the notation.

Acting on a vertex $v$, only labels of that vertex, $\mu$, and its two
neighboring ones, $\mu_{\pm}$ as well as the connecting edge labels,
$k_{\pm}$ are changed. We therefore drop all other labels in the
notation for states $|\mu_-,k_-,\mu,k_+,\mu_+\rangle$, symbolically
\begin{equation}
|\mu_-,k_-,\mu,k_+,\mu_+\rangle=\mbox{
\begin{picture}(200,15)(0,0)
 \put(0,5){\line(1,0){200}}
 \put(50,5){\circle*{5}}
 \put(100,5){\circle*{5}}
 \put(150,5){\circle*{5}}
 \put(50,-3){\makebox(0,0){$\mu_-$}}
 \put(100,-3){\makebox(0,0){$\mu$}}
 \put(150,-3){\makebox(0,0){$\mu_+$}}
 \put(25,10){\makebox(0,0){$\cdots$}}
 \put(75,12){\makebox(0,0){$k_-$}}
 \put(125,12){\makebox(0,0){$k_+$}}
 \put(175,10){\makebox(0,0){$\cdots$}}
\end{picture}}\,,
\end{equation}
 which have
connection representation
\begin{eqnarray*}
 \langle K_{\vp},A_x|\mu_-,k_-,\mu,k_+,\mu_+\rangle&:=& \exp(i\mu_-
 \gamma K_{\vp}(v_-))
 \exp\left(\case{1}{2}ik_-\smallint_{v_-}^v(A_x+\eta')\md x\right)  \exp(i\mu
 \gamma K_{\vp}(v))\\
&& \exp\left(\case{1}{2}ik_+\smallint_{v}^{v_+}
 (A_x+\eta')\md x\right) \exp(i\mu_+ \gamma K_{\vp}(v_+))\,.
\end{eqnarray*}
The action of the contributions to the Hamiltonian constraint then is,
first,
\begin{eqnarray}
\hat{H}_{\rm C}|\vec{\mu},\vec{k}\rangle &=&
\frac{\lP}{2\sqrt{2}G\gamma^{3/2}\delta^2}\left(
|\mu|\left(\sqrt{|k_++k_-+1|}-\sqrt{|k_++k_--1|}\right)\right.\\
&&\times\left(|\mu_-,k_-,\mu+2\delta,k_+,\mu_+\rangle+ 
|\mu_-,k_-,\mu-2\delta,k_+,\mu_+\rangle\right.\nonumber\\
 &&\quad-  2(1+2\gamma^2\delta^2(1-\Gamma_{\vp}^2(\vec{\mu},\vec{k})))
  |\mu_-,k_-,\mu,k_+,\mu_+\rangle)\nonumber\\
 &&\left. -4\gamma^2\delta^2\sgn_{\delta/2}(\mu)  \sqrt{|k_++k_-|}
 \Gamma_{\vp}'(\vec{\mu},\vec{k})
  |\mu_-,k_-,\mu,k_+,\mu_+\rangle\right) \nonumber\\
 && +\hat{H}_{{\rm
    matter},v}|\mu_-,k_-,\mu,k_+,\mu_+\rangle\,,\nonumber
\end{eqnarray}
where for $\Gamma_{\vp}$ and its derivative we have to insert
eigenvalues as functions of graph labels. Depending on the
quantization chosen, this may require further labels beyond $k_+$ and
$k_-$ written here explicitly. Similarly, the expression
$\sqrt{|k_++k_-+1|}-\sqrt{|k_++k_--1|}$ which occurs if $v_{\pm}$ were
not vertices of the original graph can depend on other labels if those
vertices were already present; these possibilities will be discussed
below but coefficients here do not change the main results of the
present paper. In the term containing $\Gamma_{\vp}'$ we introduced
the function
\[
 \sgn_{\delta/2}(\mu):=\frac{1}{\delta} (|\mu+\delta/2|-
|\mu-\delta/2|)= \left\{\begin{array}{cl} 1 & \mbox{ for
}\mu\geq\delta/2\\ 2\mu/\delta & \mbox{ for } -\delta/2<\mu<\delta/2\\
-1 & \mbox{ for }\mu\leq-\delta/2 \end{array}\right.
\]
as it follows from $\Delta_{\vp}\hat{V}$ and also occurs in
\begin{eqnarray}
 \hat{H}_{\rm R}|\vec{\mu},\vec{k}\rangle &=&
\frac{\lP}{4\sqrt{2}G\gamma^{3/2}\delta^2}
\sgn_{\delta/2}(\mu) \sqrt{|k_++k_-|} \\
&& \times \left(|\mu_-,k_-,\mu+\case{1}{2}\delta,k_++2,\mu_++\case{1}{2}\delta
\rangle\right.
 -|\mu_-,k_-,\mu+\case{1}{2}\delta,k_++2,\mu_+-
\case{1}{2}\delta\rangle \nonumber\\
&&+   |\mu_-,k_-,\mu-\case{1}{2}\delta,k_++2,\mu_++\case{1}{2}\delta\rangle
-   |\mu_-,k_-,\mu-\case{1}{2}\delta,k_++2,\mu_+-\case{1}{2}\delta
\rangle\nonumber\\
 &&-  |\mu_-,k_-,\mu+\case{1}{2}\delta,k_+-2,\mu_++\case{1}{2}\delta\rangle
 +   |\mu_-,k_-,\mu+\case{1}{2}\delta,k_+-2,\mu_+-\case{1}{2}\delta
\rangle\nonumber\\
 &&-   |\mu_-,k_-,\mu-\case{1}{2}\delta,k_+-2,\mu_++\case{1}{2}\delta\rangle
 +
\left.|\mu_-,k_-,\mu-\case{1}{2}\delta,k_+-2,\mu_+-\case{1}{2}\delta
\rangle\right)\nonumber
\end{eqnarray}
and analogously for $\hat{H}_{\rm L}$ where $k_-$ and $\mu_-$ change
instead of $k_+$ and $\mu_+$.

It is convenient to suppress the vertex labels and write explicitly
only changes in $k_e$ with coefficients given by vertex operators
changing only vertex labels but potentially depending on neighboring
edge labels. To that end we introduce
$\hat{C}_0(\vec{k}):=\hat{H}_{\rm C}$ together with
\begin{eqnarray}
 \hat{C}_{{\rm R}\pm}(\vec{k}) |\mu_-,\mu,\mu_+\rangle &:=& 
\pm \frac{\lP}{4\sqrt{2}G\gamma^{3/2}\delta^2}
\sgn_{\delta/2}(\mu)  \sqrt{|k_++k_-|}
 \left(|\mu_-,\mu+\case{1}{2}\delta,\mu_++\case{1}{2}\delta\rangle\right.
\nonumber\\
 &&- |\mu_-,\mu+\case{1}{2}\delta,\mu_+-\case{1}{2}\delta\rangle+
   |\mu_-,\mu-\case{1}{2}\delta,\mu_++\case{1}{2}\delta\rangle\nonumber\\
 &&- \left.
|\mu_-,\mu-\case{1}{2}\delta,\mu_+-\case{1}{2}\delta\rangle 
\right) \label{CR}\\ 
 \hat{C}_{{\rm L}\pm}(\vec{k}) |\mu_-,\mu,\mu_+\rangle &:=& 
\pm \frac{\lP}{4\sqrt{2}G\gamma^{3/2}\delta^2}
\sgn_{\delta/2}(\mu) \sqrt{|k_++k_-|}
 \left(|\mu_-+\case{1}{2}\delta,\mu+\case{1}{2}\delta,\mu_+\rangle
\right.\nonumber\\
 &&-  |\mu_--\case{1}{2}\delta,\mu+\case{1}{2}\delta,\mu_+\rangle+
   |\mu_-+\case{1}{2}\delta,\mu-\case{1}{2}\delta,\mu_+\rangle\nonumber\\
 &&- \left.  |\mu_--\case{1}{2}\delta,\mu-\case{1}{2}\delta,\mu_+\rangle\right)
 \label{CL}
\end{eqnarray}
and the constraint becomes schematically
\begin{eqnarray}
\hat{H}[N]\psi(\vec{k})
 &=&\sum_v N(v)\bigl(\hat{C}_0(\vec{k}) \psi(\ldots,k_-,k_+,\ldots)\\
&&+\hat{C}_{{\rm R}+}(\vec{k})\psi(\ldots,k_-,k_++2,\ldots)
+\hat{C}_{{\rm R}-}(\vec{k})\psi(\ldots,k_-,k_+-2,\ldots)
\nonumber\\
&&+\hat{C}_{{\rm L}+}(\vec{k}) \psi(\ldots,k_-+2,k_+,\ldots)
+\hat{C}_{{\rm L}-}(\vec{k})\psi(\ldots,k_--2,k_+,\ldots)
\bigr) \,.\nonumber
\end{eqnarray}

\subsubsection{Difference equation}

Until now, the particular choice of vertices $v_{\pm}$ only affected
the form of coefficients in $\hat{C}_0$ which will not be very
important for understanding the evolution scheme. Now, we assume that
$v_{\pm}$ have already been present in the graph before acting with
the constraint operator. The number of labels in a state then does not
change after acting and we can write the constraint equation
equivalently as a set of coupled difference equations.

Upon transforming to the triad representation by expanding
$|\psi\rangle= \sum_{\vec{k},\vec{\mu}}\psi(\vec{k}, \vec{\mu})
|\vec{k},\vec{\mu}\rangle$, the constraint equation
$\hat{H}[N]|\psi\rangle=0$ for all $N$ becomes equivalent to the set
\begin{eqnarray}\label{DiffEq}
 && ~~
 \hat{C}_{{\rm R}+}(k_-,k_+-2)^{\dagger}\psi(\ldots,k_-,k_+-2,\ldots)
 +\hat{C}_{{\rm
R}-}(k_-,k_++2)^{\dagger}\psi(\ldots,k_-,k_++2,\ldots)\nonumber\\ 
 &&+\hat{C}_{{\rm L}+}(k_--2,k_+)^{\dagger}\psi(\ldots,k_--2,k_+,\ldots)
 +\hat{C}_{{\rm L}-}(k_-+2,k_+)^{\dagger}\psi(\ldots,k_-+2,k_+,\ldots)
\nonumber\\ 
 &&+\hat{C}_0(k_-,k_+)^{\dagger}\psi(\ldots,k_-,k_+,\ldots)=0
\end{eqnarray}
of coupled difference equations, one for each vertex. The adjoints
are taken in the vertex Hilbert spaces only, i.e.\
\begin{eqnarray}
 \hat{C}_{{\rm R}\pm}(k)^{\dagger} |\mu_-,\mu,\mu_+\rangle &:=& 
\mp \frac{\lP}{4\sqrt{2}G\gamma^{3/2}\delta^2} \sqrt{|k_++k_-|}
 \left(\sgn_{\delta/2}(\mu+\case{1}{2}\delta)|\mu_-,\mu+\case{1}{2}\delta,
\mu_++\case{1}{2}\delta\rangle\right.\nonumber\\
 &&-   \sgn_{\delta/2}(\mu+\case{1}{2}\delta)|\mu_-,\mu+\case{1}{2}\delta,
\mu_+-\case{1}{2}\delta\rangle\nonumber\\
&&+
   \sgn_{\delta/2}(\mu-\case{1}{2}\delta)|\mu_-,\mu-\case{1}{2}\delta,
\mu_++\case{1}{2}\delta\rangle\nonumber\\
 &&-  \left. \sgn_{\delta/2}(\mu-\case{1}{2}\delta)
|\mu_-,\mu-\case{1}{2}\delta,
\mu_+-\case{1}{2}\delta\rangle \right)\label{CRdag}
\end{eqnarray}

So far, we used the ordering with the quantization of triad components
to the right, and the resulting operators are thus not
symmetric. After the construction we can now also consider other
orderings, most importantly the symmetric one. For this ordering, the
coefficient operator $\hat{C}_{{\rm R}_+}(k_-,k_+-2)^{\dagger}$ in
(\ref{DiffEq}) is replaced by $\frac{1}{2}(\hat{C}_{{\rm
R}_+}(k_-,k_+-2)^{\dagger}+\hat{C}_{{\rm R}_-}(k_-,k_+))$. From
(\ref{CR}) and (\ref{CRdag}) one thus obtains coefficients
proportional to $\sgn_{\delta/2}(\mu\pm\case{1}{2}\delta)\sqrt{|k_++k_--2|}+
\sgn_{\delta/2}(\mu)\sqrt{|k_++k_-|}$ for the terms multiplying
$\psi(\ldots,k_-,k_+-2,\ldots)$, the sign in $\mu\pm\delta/2$
depending on whether $\mu$ is raised or lowered in the term. It is
easy to see that these values, unlike the coefficients
$\sgn_{\delta/2}(\mu)\sqrt{|k_++k_-|}$ for the non-symmetric ordering,
are non-zero for all $k_{\pm}$ if $\mu\not=\mp\frac{1}{4}\delta$. If
$\mu=\mp\frac{1}{4}\delta$, the coefficients become zero for
$k_++k_-=1$. The coefficients in the non-symmetric ordering, on the
other hand, are zero if $k_++k_-=0$ irrespective of the value of
$\mu$. This has consequences for the singularity problem as discussed
briefly below and in more detail in
\cite{SphSymmSing}.

\subsection{Regularization issues and anomalies}

The special nature of 1-dimensional graphs relevant for spherically
symmetric spin network states makes some issues in the context of the
Hamiltonian constraint more complicated than in the full theory. For
instance, to have a well-defined operator after regulators are removed
it is necessary that the constraint operator (\ref{HOp}) annihilates
spin network states based on graphs without any vertices. Otherwise,
there would remain infinitely many contributions of the same non-zero
value after the continuum limit is performed. 

\subsubsection{Action on special vertices}

For this issue it is sufficient to consider the commutators of the
volume operator with holonomies appearing on the right hand side of
the constraint operator. While the volume operator annihilates states
without vertices, this is not obvious for the second contribution to
the commutator where we first act with a holonomy. The explicit form
of the matrix elements, however, shows that the commutators appearing
in the constraint always annihilate states with zero vertex label
$\mu_v$ such that we obtain a well-defined operator from the
non-symmetric ordering in the continuum limit. Note that the
argumentation here is less trivial than in the full theory, where one
was able to refer to planarity of vertices obtained after multiplying
with an edge holonomy in the commutator. Here, one has to use an
explicit computation of the commutator which then results in the same
conclusion.

As in the full theory, however, we do not obtain a well-defined
operator if we use the symmetric ordering before removing the
regulator. Nevertheless, one can certainly take the continuum limit of
the non-symmetric operator, which is densely defined, and symmetrize
it by adding its adjoint. This does result in a well-defined symmetric
operator, showing that there is still an ordering ambiguity, but it is
different from the symmetric ordering considered before because the
adjoint removes vertices. This remark also applies to the full theory:
regularization arguments cannot fix ordering ambiguities since a
reordering is always possible, and may even be forced for other
reasons, after the construction of operators. To fix ordering
ambiguities one needs additional arguments, such as the singularity
statements provided below.

A closer look at the behavior in vertices reveals that the situation
can be seen as related to planarity at least heuristically. For the
conclusion that the action of the Hamiltonian constraint is zero if
there is no vertex relies on the presence of $\sgn(\mu_v)$ in some
terms which would otherwise not vanish for $\mu_v=0$. (Since some of
the signs are replaced by $\sgn_{\delta/2}(\mu_v\pm\delta/2)$ in the
symmetric ordering, the continuum limit does not exist in this case.) 
The absolute values, as noted earlier, appear because of the residual
gauge transformation $\mu_v\mapsto-\mu_v$, which itself is a
consequence of isotropy in the spherical orbits. Connection components
in the $\vt$- and $\vp$-directions then merge into one local
kinematical degree of freedom, and in that sense vertices created by a
single holonomy, as they appear in commutators, can be considered
planar. This is not automatic, however, for more general vertices
since they can be interpreted as incorporating $\vp$ as well as $\vt$
information and are thus not annihilated. In polarized cylindrical
symmetry, where similar states and operators can be written, there is
no such isotropy in orbits and there are two independent vertex
labels. Planar vertices then only arise if one of the labels vanishes,
in which case they are automatically annihilated by commutators as
they always contain at least one of the two vertex labels as a factor.

Similarly, in the full theory one can use the planar nature of newly
created vertices to conclude that two Hamiltonian constraint operators
with different lapse functions commute \cite{QSDI} and then discuss
the issue of anomalies. Again, we do not have this argument at our
disposal since there is no notion of planarity in the reduced,
1-dimensional manifold. If the interpretation of planarity given above
is used, there is no simple argument since now different holonomies,
one for the vertex dependence and one for the commutator, have to be
considered. This may indicate that the argument for anomaly freedom in
the full theory is too simple, as it has indeed been criticized before
\cite{S:ClassLim,LoopCorr} on heuristic grounds of not giving local
propagating degrees of freedom, and for other reasons more recently in
\cite{NPZRev}. In this context, one-dimensional
models can be helpful because they provide two treatable situations
where local physical degrees may (polarized Einstein--Rosen) or may
not (spherical symmetry) be present (see also
\cite{SymmRed}). Nevertheless, the Hamiltonian constraint operators
appear very similar such that their properties do not obviously show
how propagating degrees of freedom would be realized.

\subsubsection{Quantization without creating new vertices}

To understand this issue it is helpful to consider different versions
of the constraint operator, and different regularization
procedures. The main question is how the new vertices $v_{\pm}$,
analogous to new edges in the full theory, are chosen. The standard
way is to create a new vertex next to $v$ not present in the original
graph. In the limit removing the regulator, $v_{\pm}$ approach $v$
which allows the expansion of edge holonomies. This is the situation
where one needs to make sure that the operator has zero action if
there is no vertex present originally because the classical constraint
regulated by a Riemann sum has contributions everywhere. It is also
the situation where, in the full theory, the argument for anomaly
freedom of \cite{AnoFree} applies.

Alternatively, $v_{\pm}$ can be chosen to be always the next vertices
already present in the original state. This would clearly prohibit the
resulting operator from being viewed as a regulated expression as
before, since $v_{\pm}$ cannot approach $v$ at a fixed state. The
viewpoint here is instead that one has an operator which reproduces
the classical expression only for suitable semiclassical states, which
still is to be proven, but not on arbitrary states (which in general
show quantum behavior or at least corrections). The first property
would be one part of the justification for using that operator, and
one condition for it to be given is that expansions of holonomies and
exponentials are valid in semiclassical regimes, which we have
demonstrated before. In addition, there must be other consistency
conditions which guarantee that the solution space to the operator is
big enough for sufficiently many semiclassical states. Here, one
usually encounters the anomaly issue: If the classical constraint
algebra is not mimicked in a certain sense by the quantum operators,
the solution space can be too strongly restricted. In what sense this
is to be ensured will have to be determined by understanding the
physics of models.

In the prescription for the constraint where $v_{\pm}$ are already
present as vertices originally, the technical issue of verifying the
commutator algebra is more involved but in addition to the fact that
the symmetric ordering is here well-defined it has the advantage that
the constraint equations can more easily be written as a system of
coupled difference equations. When $v_{\pm}$ are created as new
vertices, on the other hand, there will be new degrees of freedom
involved after each action of the constraint which makes it more
difficult to be formulated as a set of equations with a given number
of independent variables. When difference equations are available, the
anomaly issue is related to the question of whether or not all coupled
difference equations are consistent with each other, i.e.\ whether or
not a consistent recurrence scheme can be formulated to solve the
equations from given initial and boundary conditions. This turns out
to be possible \cite{SphSymmSing} despite of the fact that the absence
of anomalies is not clear yet. There are thus sufficiently many
solutions for this version of the operator.  Still, an analysis of
possible anomalies is of considerable interest as it can, for
instance, reduce the ambiguities discussed, e.g., in \cite{AlexAmbig},
but has to rely on explicit computations which will not be pursued
here.

\subsubsection{Diffeomorphism constraint}

In the preceding paragraphs we only discussed the Hamiltonian
constraint and operator equations it implies, but the constraint
algebra also has to be tested in combination with the diffeomorphism
constraint. As in the full theory \cite{ALMMT}, there is no operator
for the infinitesimal diffeomorphism constraint, but the exponentiated
version corresponding to finite diffeomorphisms can be quantized
analogously \cite{SphSymm}. It simply moves vertices implying that
their position in the background manifold $B$ has no physical
meaning. Also the coordinate position of vertices $v_{\pm}$, if they
are created by the Hamiltonian constraint, is meaningless at the
diffeomorphism invariant level, but creating a new vertex between two
given ones is still meaningful at the level of diffeomorphism classes
of graphs. The version of the constraint operator where no new
vertices are created can thus be formulated equally at the
diffeomorphism invariant or non-invariant sectors. When new vertices
are created, on the other hand, we need to choose their positions, or
work directly on diffeomorphism invariant classes without any such
choice being necessary.

The latter procedure, i.e.\ formulating Hamiltonian constraints which
create new vertices directly at the diffeomorphism invariant level can
indeed be done despite general statements to the contrary
sometimes encountered in the literature. Those statements point out
the fact that the lapse function $N$ appears in the smeared
Hamiltonian constraint and also in a quantization such as
(\ref{FullHam}). The factors $N(v)$ in vertex contributions
of the operator are not diffeomorphism invariant which renders a
simultaneous imposition of the Hamiltonian and diffeomorphism
constraints impossible.

However, after quantization each vertex contribution is a well-defined
operator and has to be imposed independently because the $N(v)$ are
free. The smeared constraint operator should be seen as a collection
of all the individual vertex constraint operators in a compact
manner. In contrast to the classical case where only the smeared
constraints have well-defined Poisson brackets, the vertex operators
have well-defined commutators as a result of the spatial discreteness
contained in a spin network state. One can thus consistently work with
the vertex contributions $\hat{H}_v$ for all $v\in B$ (which are zero
whenever $\hat{H}_v$ acts on a state which does not have $v$ as a
vertex) together with the diffeomorphism constraint. Since $N(v)$ is
no longer involved, there is no problem of defining the Hamiltonian
constraints on the diffeomorphism invariant sector.

\section{Discussion}
\label{s:Disc}

In this paper we have constructed the Hamiltonian constraint operator
for spherically symmetric models within a loop quantization and
computed its full action. Explicit calculations were facilitated by a
choice of new variables which are a mixture of connection and
extrinsic curvature components. This led to several simplifications as
compared to the pure connection variables used previously in
\cite{SphSymm,SphSymmVol,CylWaveVol,PolCylVol}, all related to the
fact that the volume operator simplifies. A volume operator with
explicitly known spectrum is also available for the pure connection
variables \cite{SphSymmVol}, but its eigenstates are not identical to
flux eigenstates. This fact implies complicated expressions for
commutators of the volume operator with holonomies which appear in the
Hamiltonian constraint. In particular the commutator
$[h_{\vp},\hat{V}]=[\cos (\frac{1}{2}A_{\vp}),\hat{V}]+
2[\sin(\frac{1}{2}A_{\vp})\Lambda_{\vp}^A(\beta), \hat{V}]$ with an
angular holonomy is cumbersome since $\sin\frac{1}{2}A_{\vp}$ has a
complicated action on volume eigenstates. Moreover, in this framework
$\beta$ does not commute with the volume operator such that there are
several different commutators in the constraint.

In the new variables introduced here, on the other hand, volume
eigenstates are identical to flux eigenstates and the relevant angle
$\eta$ does commute with the volume. Thus, there are less
commutators with different action, and each of them is easy to
compute. The relevant calculations of matrix elements of the
constraint operator are no more involved than in homogeneous cases
\cite{IsoCosmo,HomCosmo}; only the constraint equation itself is more
complicated to solve or discuss since the system involves infinitely
many kinematical degrees of freedom.

\subsection{Testing the full theory}

Symmetric models are introduced to test a possible full theory in
simpler situations and to derive physical applications in a more
direct way. For reliable results it is essential that a model is as
close to a full formulation as possible in order to ensure that there
are no artefacts from using simplifications of the
model.\footnote{This is certainly complicated in loop quantum gravity,
not the least because the specific full theory is unknown so
far. Nevertheless, if models are widely studied and general
realizations with welcome properties, such as those concerning
singularities or conceptual and phenomenological aspects, have been
identified, one should hope that a full theory can be realized
in such a manner that it reduces to those model situations in
corresponding regimes. This viewpoint puts the burden on constructing
the full theory as well as understanding its reduction. The issue is
general since one is always forced to employ approximations, whether
by reduction or otherwise, to understand physical applications of the
full theory. Without those applications, blind constructions of
possible full theories are physically empty.}  In this respect one may
worry that the introduction of new variables here spoils the relation
to the full theory since it leads to key simplifications in an
otherwise barely treatable system. Yet, the quantization in new
variables is in many respects {\em closer} to the full theory than the
previous one, which we illustrate with the following examples in the
context of this paper.

First, the volume operator is constructed immediately from flux
operators and does not contain functions of connection components
which need to be rewritten before they can be represented on the
Hilbert space. Thus, the volume operator is less ambiguous than in the
pure connection variables of \cite{SphSymmVol}, a feature shared by
both the full volume operator \cite{AreaVol,Vol2,Flux} and that of
homogeneous models \cite{CosmoII}. In this context we can also discuss
the issue of level splitting which was observed even for a single
vertex contribution of the volume operator in
\cite{SphSymm}. With the simple volume operator derived here such a
level splitting does not occur, and the vertex spectrum is identical
to that of a homogeneous model which has a single rotational axis
through each point. Level splitting then occurs only when several
vertices are considered, which explicitly brings in the inhomogeneous
properties.

Secondly, the construction of the Hamiltonian constraint operator fits
in the general scheme developed in the full theory in \cite{QSDI} and
generalized to homogeneous models in \cite{CosmoIII,HomCosmo,Spin}. In
particular the presence of possibly non-vanishing covariant components
of the spin connection in a symmetric model has to be dealt with in a
special way compared to the full theory. This at first makes it more
difficult to interpret the results since the relation to the full
theory is not as close. However, the fact that it is possible to treat
all homogeneous models in the same way strongly suggests that there is
a general procedure and results are not caused by artefacts in a
symmetric context. In this paper we have seen that this general
procedure even extends to inhomogeneous symmetric models which further
supports the whole construction. The new variables introduced here,
even though they were motivated independently by implying a simplified
volume operator, automatically implement the required subtraction of
the spin connection in homogeneous components. 

This is a very non-trivial test of the loop quantization procedure: by
using these variables both the volume operator and the Hamiltonian
constraint become closer to what is known from homogeneous models and
the full theory. Moreover, this works only when the special form of
spherically symmetric spin connections and extrinsic curvatures (or
those of the Einstein--Rosen model for which the same procedure works)
is taken into account. Otherwise, it would be far from clear that the
canonical transformation employed here to make triad components into
momenta amounts to a subtraction of homogeneous spin connection
components. The spherically symmetric model is in between homogeneous
models and the full theory, and indeed one can observe both
homogeneous and inhomogeneous aspects. The subtraction is done only in
the angular components which correspond to homogeneous directions,
while for inhomogeneous directions we still have to use the connection
component $A_x$ as would be the case in the full theory. This
automatically arises from the canonical transformation, and is
essential in obtaining a constraint operator with the correct
classical limit.

One may think that by way of our canonical transformation we go back
from Ashtekar variables to ADM like variables with extrinsic curvature
as configuration variable. This is, however, only partially true since
we do not use the $x$-component of extrinsic curvature, but the
Ashtekar connection component instead. Thus, only homogeneous
directions are affected by the transformation, while the inhomogenous
components are left unchanged. In fact, using extrinsic curvature
components throughout, which would be possible from the point of view
of the symplectic structure, would result in a different Hamiltonian
constraint operator. For instance, the dependence on $\eta$ in
(\ref{holssin}) results from insertions of $\Lambda$ in (\ref{HOp}),
not from $\eta$ in holonomies as it would happen if $A_x$ was replaced
by $K_x$.  Retaining $A_x$ as configuration variable and as an
argument of basic holonomies is crucial for an operator constructed
along the lines of the full theory. It is certainly possible to
construct constraint operators also with ADM like variables, possibly
after employing a Bohr representation as suggested recently
\cite{HWBlackHole}. But such a quantization, in contrast to one
following a general scheme for the full theory and symmetric models as
employed here, does not have contact with full quantum gravity which
makes reliable conclusions more problematic.

Moreover, in ADM variables exponentials are unmotivated in contrast to
using holonomies in theories of connections. As demonstrated by
different examples in \cite{SphSymm,AnisoPert,LivRev}, a
representation with discontinuous exponentiated connection or
extrinsic curvature components is induced in models of loop quantum
gravity by the full representation. In \cite{SphSymm} it is also shown
that in one-dimensional models one could construct many different
representations where also fluxes could be quantized by discontinuous
exponentials while the conjugate connection components have direct
actions without exponentiation. The representation is thus far from
unique if only the model is considered, and physical properties can
crucially depend on it. Only by relating models to the full theory
with its unique representation \cite{LOST,WeylRep} can a reliable
basis for using one representation be given. From this perspective, it
may be worthwhile to view the representation constructed in
\cite{HWHorizon}, which is related to that of \cite{HWBlackHole} by a
gauge fixing, also as a gauge fixing done in a loop formulation
removing the connection components $A_x$ which from our point of view
cannot be replaced by extrinsic curvature components. This could
provide an embedding of that representation in loop quantum gravity.

\subsection{Key simplifications in symmetric models}

While the general scheme discussed in the preceding subsection shows
that results can be trusted as general loop properties, rather than
properties belonging to a given model which may or may not be shared
by other models or a full theory, it also allows one to see how key
simplifications arise. They have already been exploited in physical
applications of homogeneous models, and are now also realized in
diagonal inhomogeneous models such as the spherically symmetric one or
Einstein--Rosen waves. With these models a large class of physical
situations, which already has been intensively studied classically and
with diverse quantum methods, is now accessible to explicit
computations in the loop framework. Applications span all areas of
gravitational physics from cosmology to black holes and gravitational
waves.

A technical feature shared by all these models is that the relevant
form of invariant connections and triads is diagonal, i.e.\ Lie
algebra valued components corresponding to independent directions such
as $x,\vt,\vp$ here are perpendicular in the internal direction. How
this leads to simplifications has been explained in detail in
\cite{SphSymm}. This diagonalization essentially amounts to a
reduction from SU(2) to U$(1)$ which is the reason for a simpler
volume operator. However, as seen by comparing this paper with
\cite{SphSymmVol}, this reduction in the gauge group is necessary but
not sufficient for simplifying the calculations. Also an appropriate
form of canonical variables is essential, which connects
simplifications in the volume operator with the construction of the
Hamiltonian constraint.

The required computations then are quite similar in homogeneous and
inhomogeneous models. Only the discussion and solution of the
constraint equation is, of course, more complicated in inhomogeneous
situations since more degrees of freedom are involved. But in all
these models the constraint operator takes an analogous form, and in
all models a triad representation is available. The latter feature is
the main difference to the full theory, besides the complication in
explicit calculations. Writing the constraint equation in the triad
representation is much more intuitive than in the connection
representation and has led to many results concerning the quantum
structure of classical singularities
\cite{Sing,DynIn,Essay,Closed,Scalar,ScalarLorentz,APS,HomCosmo,Spin,SphSymmSing}.
Again, one may worry that this difference to the full theory implies
special features of models which may be misleading. However, this is
not the case since we just transform an equation which we have in the
models as well as in the full theory into a different form more
suitable to applications. We are, however, not changing the equation
itself or its solutions.

In this triad representation we then obtain difference equations,
which are of different complexity: ordinary in the vacuum isotropic
case, partial in homogeneous models or even partial in infinitely many
variables in inhomogeneous models. In all cases, the nature of being a
difference equation is derived from holonomy operators appearing in
the constraint, with coefficients determined by commutators with the
volume operator. Computing those coefficients can be done with equal
ease and no new ingredients in all models, and is crucial for the
discussion of singularities (see also \cite{DegFull} for a recent
summary).

Another shared feature of the models is the relation between the
Euclidean and Lorentzian constraints, which is always simple and only
involves a rescaling of the holonomy part of the constraint. This is
in contrast to the full theory where the relation is much more
complicated and the Lorentzian constraint, so far, can only be
quantized by introducing additional commutators between the Euclidean
constraint and the volume operator \cite{QSDI}. At this point it is
not clear if the simplification in symmetric models is general enough
to lead to reliable results. This can, however, be tested directly
since the procedure of the full theory is certainly available in the
models, too. Usually, one first studies the simpler version in order
to understand the physical implications, before more complicated
choices can be compared (see, e.g., \cite{Scalar} compared to
\cite{ScalarLorentz}). So far, the more complicated version only
implies higher order equations and more involved coefficients, but no
crucial differences.

\subsection{Physical applications}

The key simplifications discussed before indicate that physical
applications can be obtained in an explicit form even at the dynamical
level. There are several features common with the formalism of
homogeneous models, such as the explicitly known matrix elements of
the Hamiltonian constraint in the Euclidean and Lorentzian form, the
fact that they are sufficiently simple functions of the spin network
labels and, crucially, the availability of a triad representation. The
latter facilitates in particular the analysis of a neighborhood of
classical singularities in order to see if they still present a
boundary to the quantum evolution. This is crucial for an
interpretation, not the formulation of models; the lack of a triad
representation in the full theory is thus not problematic for trusting
the model.

However, there are certainly also complications as compared to
homogeneous models with only finitely many kinematical degrees of
freedom. The spherically symmetric quantum constraint equation in the
triad representation is a partial difference equation in potentially
infinitely many variables and, depending on the form of the
quantization, the number of degrees of freedom involved at each time
step may not even be constant (if the constraint operator creates new
vertices). Moreover, there are many more ways to approach a classical
singularity on midisuperspace, depending also on gauge choices, and
there are different versions of singularities. This makes general
statements about their removal or persistence more complicated since
for that classical singularities first have to be located in
midisuperspace before the possibility of unique extensions of physical
wave functions is discussed. Yet, in one-dimensional models this is
possible and shows many crucial new features \cite{SphSymmSing}. The
ordering, for instance is more restricted than in homogeneous models
and a symmetric one is preferred. While homogeneous models are
non-singular for the non-symmetric and symmetric ordering because the
evolution can be continued even in cases where coefficients of the
difference equation can become zero \cite{Sing}, in inhomogeneous
models this is not possible at least for the direct operator following
from a non-symmetric version. As noted before, the symmetric ordering,
on the other hand, has generically non-zero coefficients in such a
manner that the evolution does not break down. Requiring a
non-singular evolution thus restricts the ordering choices in
inhomogeneous models.

For other physical applications it is helpful to have approximation
schemes available which allow one to isolate the essential quantum
modifications to classical equations.  In homogeneous models,
effective classical equations of the form introduced in
\cite{Inflation,Closed,Metamorph} have been essential in many recent
physical applications in the context of cosmology
\cite{InflationWMAP,Robust,BounceClosed,Oscill,BounceQualitative,LoopFluid,Cyclic,EmergentLoop,EmergentNat,NonChaos,ChaosLQC,EffHam,GenericInfl,GenericBounce,PowerLoop,PowerPert}. These
are classical equations in that they are ordinary differential
equations in coordinate time which are much easier to handle than
difference equations
\cite{DynIn,Closed,Scalar,ScalarLorentz,GenFunc,GenFuncBI,PreClassBI,ClosedExp,ContFrac,APS}. Quantum
effects are then imported by comparing the expectation value of the
quantum constraint with the classical constraint equation, which can
also be analyzed and justified by comparing the motion of quantum wave
packets with solutions to the effective equations
\cite{Time,SemiClassEmerge}. (This is part of a general procedure
which also contains the usual effective action techniques
\cite{Schilling,EffAc,Perturb,Josh}.) Similarly, effective classical
equations for inhomogeneous models, which then would be partial
differential equations in space-time coordinates, would be much easier
to deal with than a difference equation in infinitely many
variables. Moreover, mathematical expertise from geometrical analysis
would be available to arrive at general conclusions.

In cosmological models, several quantum effects have been observed
which give rise to different terms in effective classical
equations. The first and most direct one was the modification of
matter Hamiltonians at small volume \cite{QSDV,InvScale}, which makes
classically diverging energy densities finite along effective
trajectories and also played a role for the removal of cosmological
singularities. The same modification is available in a spherically
symmetric model with matter; but since here even a vacuum model would
have a classical (Schwarzschild) singularity, it cannot be expected to
be sufficient for a non-singular effective formulation.

Other modifications come from the gravitational part of the constraint
and are thus available even in the vacuum case. There are perturbative
corrections which can be interpreted as being analogous to higher
curvature terms, and non-perturbative ones in the case of a
non-vanishing spin connection. (The latter can be viewed as providing
a natural cut-off on intrinsic curvature, analogously to the extrinsic
curvature cut-off from a quantum matter Hamiltonian in a homogeneous
situation.) The non-perturbative modifications have mainly been used
so far in the Bianchi IX model where they remove the classical chaos
\cite{NonChaos,ChaosLQC}. They are derived by quantizing the spin
connection potential term of the Hamiltonian constraint in accordance
with the general scheme described above. Since the spin connection
contains inverse powers of the triad components, it will obtain
modifications from the quantization of inverse powers
\cite{QSDV,InvScale} similarly to a matter Hamiltonian.

Here, the same procedure is available where we also have an intrinsic
curvature potential with the spin connection depending on inverse
powers of $E^{\vp}$. Using eigenvalues of the quantized potential in
the classical constraint equation then provides effective classical
equations showing one main quantum effect from the cut-off of
intrinsic curvature. Comparison with the Schwarzschild solution shows
that this indeed provides modifications at the right place, namely
close to the classical singularity where $E^{\vp}$ is small. In
asymptotic regimes or around the horizon of massive black holes, on
the other hand, $E^{\vp}$ is large and so the behavior there remains
classical. This would be different had we used other metric variables
such as the co-triad or metric instead of the densitized triad as,
e.g., in \cite{HWBlackHole}. Then the component whose inverse appears
in the spin connection (\ref{GammaK}) would be $e_x$, which for
Schwarzschild is large at the horizon but also at the classical
singularity and so there would be no quantum corrections there. In
asymptotic regimes, on the other hand, $e_x$ would be of the order one:
modifications would be noticable but completely
unwanted. Thus, similarly to homogeneous models \cite{HomCosmo} we see
that the issue of singularity removal will crucially depend on the
canonical variables used. Densitized triad components, which we have
to use anyway since they are part of the basis of the full background
independent quantization, are well-suited for this aim. These
statements are only preliminary as of now because no consistent set of
effective equations for inhomogeneous cases has been derived so far.

Besides singularities, horizons would be the second interesting aspect
to be studied. As just noted, we do not expect strong quantum
modifications there at least for massive black holes. But there are
other interesting aspects which are usually related to quantum models,
such as the issue of black hole entropy, horizon degrees of freedom
and fluctuations, and Hawking radiation. All these issues, for
instance in the context of the scenario of \cite{BHPara}, require
solving the Hamiltonian constraint which can provide important
feedback on the viability of a given quantization scheme, and thus
test the full theory and restrict quantization ambiguities. The
isolated or dynamical horizon framework
\cite{IH,IHApp,IHPhase,DynHorLett,DynHor,HorRev} provides an ideal
setting for an analysis at the classical and quantum levels. If a
horizon is isolated or slowly evolving \cite{SlowHor}, certain terms
in the quantum constraint become negligible such that the constraint
simplifies \cite{Horizon}. This provides an approximation scheme to
understand the physics at the horizon or, in a perturbative form,
around the horizon. Also dynamical processes can then be studied in a
controlled manner because not just isolated but also slowly evolving
horizons are allowed.  In this way one can derive physical information
about black hole or other horizons, but also study general issues of
the Hamiltonian constraint such as observables.

\section*{Acknowledgements}

We thank Kevin Vandersloot for discussions and for checking some of
the formulas.


\begin{thebibliography}{100}

\bibitem{Rov:Loops}
C.\ Rovelli,
\newblock Loop Quantum Gravity,
\newblock {\em Living Rev.\ Rel.} 1 (1998) 1, [gr-qc/9710008],
\newblock {\tt http://www.livingreviews.org/Articles/Volume1/1998-1rovelli}; 
{\em Quantum Gravity} (Cambridge University Press, Cambridge, 2004)

\bibitem{ThomasRev}
T.\ Thiemann,
\newblock Introduction to Modern Canonical Quantum General Relativity,
  [gr-qc/0110034]

\bibitem{ALRev}
A.\ Ashtekar and J.\ Lewandowski,
\newblock Background independent quantum gravity: A status report,
\newblock {\em Class.\ Quantum Grav.} 21 (2004) R53--R152, [gr-qc/0404018]

\bibitem{SymmRed}
M.\ Bojowald and H.~A.\ Kastrup,
\newblock Symmetry Reduction for Quantized Diffeomorphism Invariant Theories of
  Connections,
\newblock {\em Class.\ Quantum Grav.} 17 (2000) 3009--3043, [hep-th/9907042]

\bibitem{CosmoI}
M.\ Bojowald,
\newblock Loop Quantum Cosmology: I. Kinematics,
\newblock {\em Class.\ Quantum Grav.} 17 (2000) 1489--1508, [gr-qc/9910103]

\bibitem{IsoCosmo}
M.\ Bojowald,
\newblock Isotropic Loop Quantum Cosmology,
\newblock {\em Class.\ Quantum Grav.} 19 (2002) 2717--2741, [gr-qc/0202077]

\bibitem{Bohr}
A.\ Ashtekar, M.\ Bojowald, and J.\ Lewandowski,
\newblock Mathematical structure of loop quantum cosmology,
\newblock {\em Adv.\ Theor.\ Math.\ Phys.} 7 (2003) 233--268, [gr-qc/0304074]

\bibitem{HomCosmo}
M.\ Bojowald,
\newblock Homogeneous loop quantum cosmology,
\newblock {\em Class.\ Quantum Grav.} 20 (2003) 2595--2615, [gr-qc/0303073]

\bibitem{Spin}
M.\ Bojowald, G.\ Date, and K.\ Vandersloot,
\newblock Homogeneous loop quantum cosmology: The role of the spin connection,
\newblock {\em Class.\ Quantum Grav.} 21 (2004) 1253--1278, [gr-qc/0311004]

\bibitem{LoopCosRev}
M.\ Bojowald and H.~A.\ Morales-T\'ecotl,
\newblock Cosmological applications of loop quantum gravity,
\newblock In: Proceedings of the Fifth Mexican School (DGFM): The Early
  Universe and Observational Cosmology,
{\em Lect.\ Notes Phys.} 646 (2004) 421--462 (Springer-Verlag), [gr-qc/0306008]

\bibitem{WS:MB}
M.\ Bojowald,
\newblock Loop quantum cosmology, In: {\em 100 Years of Relativity -- Space-Time Structure: Einstein and Beyond}, Ed.: A.~Ashtekar,
\newblock World Scientific, Singapore, 2005, [gr-qc/0505057]

\bibitem{LivRev}
M.\ Bojowald,
\newblock Loop Quantum Cosmology,
\newblock {\em Living Rev.\ Relativity}, to appear

\bibitem{SemiClass}
M.\ Bojowald,
\newblock The Semiclassical Limit of Loop Quantum Cosmology,
\newblock {\em Class.\ Quantum Grav.} 18 (2001) L109--L116, [gr-qc/0105113]

\bibitem{Time}
M.\ Bojowald, P.\ Singh, and A.\ Skirzewski,
\newblock Coordinate time dependence in quantum gravity,
\newblock {\em Phys.\ Rev.\ D} 70 (2004) 124022, [gr-qc/0408094]

\bibitem{SemiClassEmerge}
P.\ Singh and K.\ Vandersloot,
\newblock Semi-classical States, Effective Dynamics and Classical Emergence in
  Loop Quantum Cosmology,
\newblock {\em Phys.\ Rev.\ D} 72 (2005) 084004, [gr-qc/0507029]

\bibitem{APS}
A.\ Ashtekar, T.\ Pawlowski, and P.\ Singh, in preparation

\bibitem{Sing}
M.\ Bojowald,
\newblock Absence of a Singularity in Loop Quantum Cosmology,
\newblock {\em Phys.\ Rev.\ Lett.} 86 (2001) 5227--5230, [gr-qc/0102069]

\bibitem{DynIn}
M.\ Bojowald,
\newblock Dynamical Initial Conditions in Quantum Cosmology,
\newblock {\em Phys.\ Rev.\ Lett.} 87 (2001) 121301, [gr-qc/0104072]

\bibitem{Inflation}
M.\ Bojowald,
\newblock Inflation from quantum geometry,
\newblock {\em Phys.\ Rev.\ Lett.} 89 (2002) 261301, [gr-qc/0206054]

\bibitem{InflationWMAP}
S.\ Tsujikawa, P.\ Singh, and R.\ Maartens,
\newblock Loop quantum gravity effects on inflation and the CMB,
\newblock {\em Class.\ Quantum Grav.} 21 (2004) 5767--5775, [astro-ph/0311015]

\bibitem{PowerLoop}
G.~M.\ Hossain,
\newblock Primordial Density Perturbation in Effective Loop Quantum Cosmology,
\newblock {\em Class.\ Quantum Grav.} 22 (2005) 2511--2532, [gr-qc/0411012]

\bibitem{NonChaos}
M.\ Bojowald and G.\ Date,
\newblock Quantum suppression of the generic chaotic behavior close to
  cosmological singularities,
\newblock {\em Phys.\ Rev.\ Lett.} 92 (2004) 071302, [gr-qc/0311003]

\bibitem{BounceClosed}
P.\ Singh and A.\ Toporensky,
\newblock Big Crunch Avoidance in ${\rm k}=1$ Semi-Classical Loop Quantum
  Cosmology,
\newblock {\em Phys.\ Rev.\ D} 69 (2004) 104008, [gr-qc/0312110]

\bibitem{Oscill}
J.~E.\ Lidsey, D.~J.\ Mulryne, N.~J.\ Nunes, and R.\ Tavakol,
\newblock Oscillatory Universes in Loop Quantum Cosmology and Initial
  Conditions for Inflation,
\newblock {\em Phys.\ Rev.\ D} 70 (2004) 063521, [gr-qc/0406042]

\bibitem{EmergentLoop}
D.~J.\ Mulryne, R.\ Tavakol, J.~E.\ Lidsey, and G.~F.~R.\ Ellis,
\newblock An emergent universe from a loop,
\newblock {\em Phys.\ Rev.\ D} 71 (2005) 123512, [astro-ph/0502589]

\bibitem{Collapse}
M.\ Bojowald, R.\ Goswami, R.\ Maartens, and P.\ Singh,
\newblock A black hole mass threshold from non-singular quantum gravitational
  collapse,
\newblock {\em Phys.\ Rev.\ Lett.} 95 (2005) 091302, [gr-qc/0503041]

\bibitem{BHPara}
A.\ Ashtekar and M.\ Bojowald,
\newblock Black hole evaporation: A paradigm,
\newblock {\em Class.\ Quantum Grav.} 22 (2005) 3349--3362, [gr-qc/0504029]

\bibitem{SphKl1}
T.\ Thiemann and H.~A.\ Kastrup,
\newblock Canonical Quantization of Spherically Symmetric Gravity in Ashtekar's
  Self-Dual Representation,
\newblock {\em Nucl.\ Phys.\ B} 399 (1993) 211--258, [gr-qc/9310012]

\bibitem{SphKl2}
H.~A.\ Kastrup and T.\ Thiemann,
\newblock Spherically Symmetric Gravity as a Completely Integrable System,
\newblock {\em Nucl.\ Phys.\ B} 425 (1994) 665--686, [gr-qc/9401032]

\bibitem{Kuchar}
K.~V.\ Kucha\v{r},
\newblock Geometrodynamics of Schwarzschild Black Holes,
\newblock {\em Phys.\ Rev.\ D} 50 (1994) 3961--3981

\bibitem{SphSymm}
M.\ Bojowald,
\newblock Spherically Symmetric Quantum Geometry: States and Basic Operators,
\newblock {\em Class.\ Quantum Grav.} 21 (2004) 3733--3753, [gr-qc/0407017]

\bibitem{SphSymmVol}
M.\ Bojowald and R.\ Swiderski,
\newblock The Volume Operator in Spherically Symmetric Quantum Geometry,
\newblock {\em Class.\ Quantum Grav.} 21 (2004) 4881--4900, [gr-qc/0407018]

\bibitem{AreaVol}
C.\ Rovelli and L.\ Smolin,
\newblock Discreteness of Area and Volume in Quantum Gravity,
\newblock {\em Nucl.\ Phys.\ B} 442 (1995) 593--619, [gr-qc/9411005],
\newblock Erratum: {\em Nucl.\ Phys.\ B} 456 (1995) 753

\bibitem{Vol2}
A.\ Ashtekar and J.\ Lewandowski,
\newblock Quantum Theory of Geometry II: Volume Operators,
\newblock {\em Adv.\ Theor.\ Math.\ Phys.} 1 (1997) 388--429, [gr-qc/9711031]

\bibitem{RecTh}
R.\ De~Pietri and C.\ Rovelli,
\newblock Geometry Eigenvalues and the Scalar Product from Recoupling Theory in
  Loop Quantum Gravity,
\newblock {\em Phys.\ Rev.\ D} 54 (1996) 2664--2690

\bibitem{Vol}
T.\ Thiemann,
\newblock Closed Formula for the Matrix Elements of the Volume Operator in
  Canonical Quantum Gravity,
\newblock {\em J.\ Math.\ Phys.} 39 (1998) 3347--3371, [gr-qc/9606091]

\bibitem{VolNum}
J.\ Brunnemann and T.\ Thiemann,
\newblock Simplification of the Spectral Analysis of the Volume Operator in
  Loop Quantum Gravity, [gr-qc/0405060]

\bibitem{QSDI}
T.\ Thiemann,
\newblock Quantum Spin Dynamics {(QSD)},
\newblock {\em Class.\ Quantum Grav.} 15 (1998) 839--873, [gr-qc/9606089]

\bibitem{CylWaveVol}
D.~E.\ Neville,
\newblock The volume operator for spin networks with planar or cylindrical
  symmetry, [gr-qc/0511005]

\bibitem{PolCylVol}
D.~E.\ Neville,
\newblock The volume operator for singly polarized gravity waves with planar or
  cylindrical symmetry, [gr-qc/0511006]

\bibitem{AshVar}
A.\ Ashtekar,
\newblock New Hamiltonian Formulation of General Relativity,
\newblock {\em Phys.\ Rev.\ D} 36 (1987) 1587--1602

\bibitem{AshVarReell}
J.~F.\ Barbero~G.,
\newblock Real Ashtekar Variables for Lorentzian Signature Space-Times,
\newblock {\em Phys.\ Rev.\ D} 51 (1995) 5507--5510, [gr-qc/9410014]

\bibitem{Immirzi}
G.\ Immirzi,
\newblock Real and Complex Connections for Canonical Gravity,
\newblock {\em Class.\ Quantum Grav.} 14 (1997) L177--L181

\bibitem{FundamentalDisc}
M.\ Bojowald and G.\ Date,
\newblock Consistency conditions for fundamentally discrete theories,
\newblock {\em Class.\ Quantum Grav.} 21 (2004) 121--143, [gr-qc/0307083]

\bibitem{BHInt}
A.\ Ashtekar and M.\ Bojowald,
\newblock Quantum Geometry and the Schwarzschild Singularity, [gr-qc/0509075]

\bibitem{DegFull}
M.\ Bojowald,
\newblock Degenerate Configurations, Singularities and the Non-Abelian Nature
  of Loop Quantum Gravity, [gr-qc/0508118]

\bibitem{FermionHiggs}
T.\ Thiemann,
\newblock Kinematical Hilbert Spaces for Fermionic and Higgs Quantum Field
  Theories,
\newblock {\em Class.\ Quantum Grav.} 15 (1998) 1487--1512, [gr-qc/9705021]

\bibitem{ScalarBohr}
A.\ Ashtekar, J.\ Lewandowski, and H.\ Sahlmann,
\newblock Polymer and Fock representations for a Scalar field,
\newblock {\em Class.\ Quantum Grav.} 20 (2003) L11--L21, [gr-qc/0211012]

\bibitem{EffHam}
G.\ Date and G.~M.\ Hossain,
\newblock Effective Hamiltonian for Isotropic Loop Quantum Cosmology,
\newblock {\em Class.\ Quantum Grav.} 21 (2004) 4941--4953, [gr-qc/0407073]

\bibitem{DiscCorr}
K.\ Banerjee and G.\ Date,
\newblock Discreteness Corrections to the Effective Hamiltonian of Isotropic
  Loop Quantum Cosmology,
\newblock {\em Class.\ Quant.\ Grav.} 22 (2005) 2017--2033, [gr-qc/0501102]

\bibitem{AmbigConstr}
K.\ Vandersloot,
\newblock On the Hamiltonian Constraint of Loop Quantum Cosmology,
\newblock {\em Phys.\ Rev.\ D} 71 (2005) 103506, [gr-qc/0502082]

\bibitem{Perturb}
A.\ Ashtekar, M.\ Bojowald, and J.\ Willis, in preparation

\bibitem{AnisoPert}
M.\ Bojowald, H.~H.\ Hern\'andez, and H.~A.\ Morales-T\'ecotl,
\newblock Perturbative degrees of freedom in loop quantum gravity:
  Anisotropies, [gr-qc/0511058]

\bibitem{QSDV}
T.\ Thiemann,
\newblock {QSD V}: Quantum Gravity as the Natural Regulator of Matter Quantum
  Field Theories,
\newblock {\em Class.\ Quantum Grav.} 15 (1998) 1281--1314, [gr-qc/9705019]

\bibitem{InvScale}
M.\ Bojowald,
\newblock Inverse Scale Factor in Isotropic Quantum Geometry,
\newblock {\em Phys.\ Rev.\ D} 64 (2001) 084018, [gr-qc/0105067]

\bibitem{CosmoIII}
M.\ Bojowald,
\newblock Loop Quantum Cosmology III: Wheeler-DeWitt Operators,
\newblock {\em Class.\ Quantum Grav.} 18 (2001) 1055--1070, [gr-qc/0008052]

\bibitem{Ambig}
M.\ Bojowald,
\newblock Quantization ambiguities in isotropic quantum geometry,
\newblock {\em Class.\ Quantum Grav.} 19 (2002) 5113--5130, [gr-qc/0206053]

\bibitem{ICGC}
M.\ Bojowald,
\newblock Loop Quantum Cosmology: Recent Progress,
\newblock In: Proceedings of the International Conference on Gravitation
  and Cosmology (ICGC 2004), Cochin, India,
  {\em Pramana} 63 (2004) 765--776, [gr-qc/0402053]

\bibitem{SphSymmSing}
M.\ Bojowald,
\newblock Non-singular black holes and degrees of freedom in quantum gravity,
\newblock {\em Phys.\ Rev.\ Lett.} 95 (2005) 061301, [gr-qc/0506128]

\bibitem{S:ClassLim}
L.\ Smolin,
\newblock The Classical Limit and the Form of the Hamiltonian Constraint in
  Non-Perturbative Quantum General Relativity, [gr-qc/9609034]

\bibitem{LoopCorr}
D.~E.\ Neville,
\newblock Long range correlations in quantum gravity,
\newblock {\em Phys.\ Rev.\ D} 59 (1999) 044032, [gr-qc/9803066]

\bibitem{NPZRev}
H.\ Nicolai, K.\ Peeters, and M.\ Zamaklar,
\newblock Loop quantum gravity: an outside view,
\newblock {\em Class.\ Quantum Grav.} 22 (2005) R193--R247, [hep-th/0501114]

\bibitem{AnoFree}
T.\ Thiemann,
\newblock Anomaly-Free Formulation of Non-Perturbative,
  Four-Dimensional Lorentzian Quantum Gravity,
\newblock {\em Phys.\ Lett.\ B} 380 (1996) 257--264, [gr-qc/9606088]

\bibitem{AlexAmbig}
A.\ Perez,
\newblock On the regularization ambiguities in loop quantum gravity,
  [gr-qc/0509118]

\bibitem{ALMMT}
A.\ Ashtekar, J.\ Lewandowski, D.\ Marolf, J.\ Mour\~ao, and T.\ Thiemann,
\newblock Quantization of Diffeomorphism Invariant Theories of Connections with
  Local Degrees of Freedom,
\newblock {\em J.\ Math.\ Phys.} 36 (1995) 6456--6493, [gr-qc/9504018]

\bibitem{Flux}
K.\ Giesel and T.\ Thiemann,
\newblock Consistency Check on Volume and Triad Operator Quantisation in Loop
  Quantum Gravity I, [gr-qc/0507036]

\bibitem{CosmoII}
M.\ Bojowald,
\newblock Loop Quantum Cosmology: II. Volume Operators,
\newblock {\em Class.\ Quantum Grav.} 17 (2000) 1509--1526, [gr-qc/9910104]

\bibitem{HWBlackHole}
V.\ Husain and O.\ Winkler,
\newblock Quantum resolution of black hole singularities,
\newblock {\em Class.\ Quantum Grav.} 22 (2005) L127--L133, [gr-qc/0410125]

\bibitem{LOST}
J.\ Lewandowski, A.\ Oko\l\'ow, H.\ Sahlmann, and T.\ Thiemann,
\newblock Uniqueness of diffeomorphism invariant states on holonomy-flux
  algebras, [gr-qc/0504147]

\bibitem{WeylRep}
C.\ Fleischhack,
\newblock Representations of the Weyl Algebra in Quantum Geometry,
  [math-ph/0407006]

\bibitem{HWHorizon}
V.\ Husain and O.\ Winkler,
\newblock Quantum black holes from null expansion operators,
\newblock {\em Class.\ Quantum Grav.} 22 (2005) L135--L141, [gr-qc/0412039]

\bibitem{Essay}
M.\ Bojowald,
\newblock Initial Conditions for a Universe,
\newblock {\em Gen.\ Rel.\ Grav.} 35 (2003) 1877--1883, [gr-qc/0305069]

\bibitem{Closed}
M.\ Bojowald and K.\ Vandersloot,
\newblock Loop quantum cosmology, boundary proposals, and inflation,
\newblock {\em Phys.\ Rev.\ D} 67 (2003) 124023, [gr-qc/0303072]

\bibitem{Scalar}
M.\ Bojowald and F.\ Hinterleitner,
\newblock Isotropic loop quantum cosmology with matter,
\newblock {\em Phys.\ Rev.\ D} 66 (2002) 104003, [gr-qc/0207038]

\bibitem{ScalarLorentz}
F.\ Hinterleitner and S.\ Major,
\newblock Isotropic Loop Quantum Cosmology with Matter II: The Lorentzian
  Constraint,
\newblock {\em Phys.\ Rev.\ D} 68 (2003) 124023, [gr-qc/0309035]

\bibitem{Metamorph}
P.\ Singh,
\newblock Effective State Metamorphosis in Semi-Classical Loop Quantum
  Cosmology,
\newblock {\em Class.\ Quantum Grav.} 22 (2005) 4203--4216, [gr-qc/0502086]

\bibitem{Robust}
M.\ Bojowald, J.~E.\ Lidsey, D.~J.\ Mulryne, P.\ Singh, and R.\ Tavakol,
\newblock Inflationary Cosmology and Quantization Ambiguities in Semi-Classical
  Loop Quantum Gravity,
\newblock {\em Phys.\ Rev.\ D} 70 (2004) 043530, [gr-qc/0403106]

\bibitem{BounceQualitative}
G.~V.\ Vereshchagin,
\newblock Qualitative Approach to Semi-Classical Loop Quantum Cosmology,
\newblock {\em JCAP} 07 (2004) 013, [gr-qc/0406108]

\bibitem{LoopFluid}
N.~J.\ Nunes,
\newblock Inflation: A graceful entrance from Loop Quantum Cosmology, 
{\em Phys.\ Rev.\ D} 72 (2005) 103510,
  [astro-ph/0507683]

\bibitem{Cyclic}
M.\ Bojowald, R.\ Maartens, and P.\ Singh,
\newblock Loop Quantum Gravity and the Cyclic Universe,
\newblock {\em Phys.\ Rev.\ D} 70 (2004) 083517, [hep-th/0407115]

\bibitem{EmergentNat}
M.\ Bojowald,
\newblock Original Questions,
\newblock {\em Nature} 436 (2005) 920--921

\bibitem{ChaosLQC}
M.\ Bojowald, G.\ Date, and G.~M.\ Hossain,
\newblock The Bianchi IX model in loop quantum cosmology,
\newblock {\em Class.\ Quantum Grav.} 21 (2004) 3541--3569, [gr-qc/0404039]

\bibitem{GenericInfl}
G.\ Date and G.~M.\ Hossain,
\newblock Genericity of inflation in isotropic loop quantum cosmology,
\newblock {\em Phys.\ Rev.\ Lett.} 94 (2005) 011301, [gr-qc/0407069]

\bibitem{GenericBounce}
G.\ Date and G.~M.\ Hossain,
\newblock Genericity of Big Bounce in isotropic loop quantum cosmology,
\newblock {\em Phys.\ Rev.\ Lett.} 94 (2005) 011302, [gr-qc/0407074]

\bibitem{PowerPert}
S.\ Hofmann and O.\ Winkler,
\newblock The Spectrum of Fluctuations in Inflationary Quantum Cosmology,
  [astro-ph/0411124]

\bibitem{GenFunc}
D.\ Cartin, G.\ Khanna, and M.\ Bojowald,
\newblock Generating function techniques for loop quantum cosmology,
\newblock {\em Class.\ Quantum Grav.} 21 (2004) 4495--4509, [gr-qc/0405126]

\bibitem{GenFuncBI}
D.\ Cartin and G.\ Khanna,
\newblock Absence of pre-classical solutions in Bianchi I loop quantum
  cosmology,
\newblock {\em Phys.\ Rev.\ Lett.} 94 (2005) 111302, [gr-qc/0501016]

\bibitem{PreClassBI}
G.\ Date,
\newblock Pre-classical solutions of the vacuum Bianchi I loop quantum
  cosmology,
\newblock {\em Phys.\ Rev.\ D} 72 (2005) 067301, [gr-qc/0505030]

\bibitem{ClosedExp}
D.\ Green and W.\ Unruh,
\newblock Difficulties with Closed Isotropic Loop Quantum Cosmology,
\newblock {\em Phys.\ Rev.\ D} 70 (2004) 103502, [gr-qc/0408074]

\bibitem{ContFrac}
M.\ Bojowald and A.\ Rej,
\newblock Asymptotic Properties of Difference Equations for Isotropic Loop
  Quantum Cosmology,
\newblock {\em Class.\ Quantum Grav.} 22 (2005) 3399--3420, [gr-qc/0504100]

\bibitem{Schilling}
A.\ Ashtekar and T.~A.\ Schilling,
\newblock Geometrical Formulation of Quantum Mechanics, In: 
{\em On Einstein's Path: Essays in Honor of Engelbert Sch\"ucking}, 
Ed.: A.\ Harvey, pages 23--65,
\newblock (Springer, New York, 1999), [gr-qc/9706069]

\bibitem{EffAc}
M.\ Bojowald and A.\ Skirzewski,
\newblock Effective Equations of Motion for Quantum Systems,
\newblock [math--ph/0511043]

\bibitem{Josh}
J.\ Willis,
\newblock {\em On the Low-Energy Ramifications and a Mathematical Extension of
  Loop Quantum Gravity},
\newblock PhD thesis, The Pennsylvania State University, 2004

\bibitem{IH}
A.\ Ashtekar, C.\ Beetle, and S.\ Fairhurst,
\newblock Isolated Horizons: A Generalization of Black Hole Mechanics,
\newblock {\em Class.\ Quantum Grav.} 16 (1999) L1--L7, [gr-qc/9812065]

\bibitem{IHApp}
A.\ Ashtekar, C.\ Beetle, O.\ Dreyer, S.\ Fairhurst, B.\ Krishnan, J.\
  Lewandowski, and J.\ Wi\'sniewski,
\newblock Isolated Horizons and their Applications,
\newblock {\em Phys.\ Rev.\ Lett.} 85 (2000) 3564--3567, [gr-qc/0006006]

\bibitem{IHPhase}
A.\ Ashtekar, A.\ Corichi, and K.\ Krasnov,
\newblock Isolated Horizons: the Classical Phase Space,
\newblock {\em Adv.\ Theor.\ Math.\ Phys.} 3 (2000) 419--478, [gr-qc/9905089]

\bibitem{DynHorLett}
A.\ Ashtekar and B.\ Krishnan,
\newblock Dynamical Horizons: Energy, Angular Momentum, Fluxes and Balance
  Laws,
\newblock {\em Phys.\ Rev.\ Lett.} 89 (2002) 261101, [gr-qc/0207080]

\bibitem{DynHor}
A.\ Ashtekar and B.\ Krishnan,
\newblock Dynamical Horizons and their Properties,
\newblock {\em Phys.\ Rev.\ D} 68 (2003) 104030, [gr-qc/0308033]

\bibitem{HorRev}
A.\ Ashtekar and B.\ Krishnan,
\newblock Isolated and dynamical horizons and their applications,
\newblock {\em Living Rev.\ Rel.} 7 (2004) 1--77, [gr-qc/0407042]

\bibitem{SlowHor}
I.\ Booth and S.\ Fairhurst,
\newblock The first law for slowly evolving horizons,
\newblock {\em Phys.\ Rev.\ Lett.} 92 (2004) 011102, [gr-qc/0307087]

\bibitem{Horizon}
M.\ Bojowald and R.\ Swiderski,
\newblock Spherically Symmetric Quantum Horizons,
\newblock {\em Phys.\ Rev.\ D} 71 (2005) 081501(R), [gr-qc/0410147]

\end{thebibliography}

\end{document}